\documentclass[11pt, romanappendices]{article}
\usepackage[margin=1in]{geometry}

\usepackage{mdframed}
\usepackage{cite}
\usepackage{stfloats}
\usepackage{relsize}
\usepackage{authblk}
\usepackage{setspace}
\usepackage{amsopn}
\usepackage{float}
\restylefloat{table}
\usepackage{amsthm}
\usepackage{algpseudocode,algorithm,algorithmicx}
\usepackage{amsxtra}
\usepackage{amsmath}
\usepackage{graphicx}
\usepackage{latexsym}
\usepackage{amsfonts}
\usepackage{amssymb}
\usepackage{mathrsfs}
\usepackage{bm}
\usepackage{yhmath}
\usepackage{verbatim}
\usepackage{float}
\usepackage{xr-hyper}
\usepackage{multirow}
\usepackage{subfloat}

\usepackage{accents}

\usepackage{epstopdf}

\usepackage[font=small,labelfont=bf]{caption}
\usepackage{subcaption}

\theoremstyle{definition}

\theoremstyle{condition}

\DeclareMathAlphabet{\bi}{OML}{cmm}{b}{it}
\DeclareMathAlphabet{\bcal}{OMS}{cmsy}{b}{n}
\DeclareMathAlphabet{\brmn}{OT1}{cmr}{bx}{n}

\DeclareMathSymbol{\R}{\mathalpha}{AMSb}{"52}

\newcommand{\bgamma}{\boldsymbol{\gamma}}

\newcommand{\brho}{\boldsymbol{\rho}}

\newcommand{\Jc}{\mathcal{J}}

\newcommand{\Lc}{\mathcal{L}}

\newcommand{\Cb}{\mathbb{C}}

\newcommand{\Rb}{\mathbb{R}}

\renewcommand{\o}{\omega}
\usepackage{amsfonts}

\graphicspath{{./images/}}

\DeclareMathAlphabet{\bi}{OML}{cmm}{b}{it}
\DeclareMathAlphabet{\bcal}{OMS}{cmsy}{b}{n}
\DeclareMathAlphabet{\brmn}{OT1}{cmr}{bx}{n}

\def \x{\mathbf{x}}

\def \y{\mathbf{y}}

\def \h{\mathbf{h}}

\def \b{\mathbf{b}}
\def \w{\mathbf{w}}

\def \d{\mathbf{d}}

\def \R{\mathbf{R}}

\def \W{\mathbf{W}}

\def \P{\mathbf{P}}

\def \F{\mathbf{F}}

\def \0{\mathbf{0}}

\begin{document}

%\supertitle{Research Journal Paper}
\title{Deep Learning for Waveform Estimation and Imaging in Passive Radar}

%\author{\au{Bariscan Yonel$^{1}$}, \au{Eric Mason$^{2}$}, \au{Birsen Yazici$^{3\corr}$}}

\author[$\dagger$]{Bariscan Yonel}
\author[$\star$]{Eric Mason}
\author[$\dagger$]{Birsen Yazici}

%\address{\add{1}{Department of Electrical, Computer and System Engineering, Rensselaer Polytechnic
%Institute, Troy, NY 12180 USA}
%\add{2}{Tactical Electronic Warfare Division, Naval Research Laboratory, Washington, DC 20375, USA}
%\add{3}{Department of Electrical, Computer and System Engineering, Rensselaer Polytechnic
%Institute, Troy, NY 12180 USA}
%%\add{4}{Current affiliation: Fourth Department, Fourth University, Address, Country Name}

\affil[$\dagger$]{Department of Electrical, Computer and System Engineering, Rensselaer Polytechnic Institute, Troy, NY 12180 USA}
\affil[$\star$]{Tactical Electronic Warfare Division, Naval Research Laboratory, Washington, DC 20375, USA}
%\email{yazici@ecse.rpi.edu}}

\date{\vspace{-5ex}}
\maketitle

\begin{abstract}
{
We consider a bistatic configuration with a stationary transmitter transmitting unknown waveforms of opportunity and a moving receiver and present a Deep Learning (DL) framework for passive synthetic aperture radar (SAR) imaging. Existing passive radar methods require two or more antennas that are either spatially separated or colocated with sufficient directivity to estimate the underlying waveform prior to imaging.
Our approach to passive radar only requires a single receiver, hence reducing cost and increasing versatility.
 %compared to currently used methods.
%Moreover, DL-based formulation is a joint estimation approach rather than a pre-processing approach, which does not require a direct path signal or de-noising of received signal.
We approach DL from an optimization perspective and formulate image reconstruction as a machine learning task.
By unfolding the iterations of a proximal gradient descent algorithm, we construct a deep recurrent neural network (RNN) that is parameterized by transmitted waveforms.
%or which a deep recurrent neural network (RNN) can be constructed by unfolding the iterations of a proximal gradient descent algorithm.
%The network becomes an inverse solver that is parameterized by the transmitted waveform for the SAR imaging problem.
We cascade the RNN structure with a decoder stage %as previously proposed in \cite{yonel2017deep, mason2017deep2, mason2017deep}
to form a recurrent-auto encoder architecture.
We then use backpropagation to learn transmitted waveforms by training the network in an unsupervised manner using SAR measurements. %obtained from the same imaging geometry.
The highly non-convex problem of backpropagation is guided to a feasible solution over the parameter space by initializing the network with the known components of the SAR forward model.
Moreover, prior information regarding the waveform structure is incorporated during initialization and backpropagation.
We demonstrate the effectiveness of the DL-based approach through extensive numerical simulations that show focused, high contrast imagery using a single receiver antenna at realistic SNR levels. 

}
\end{abstract}

\maketitle

\section{Introduction}\label{sec1}

\subsection{Motivations}

Deep Learning (DL) has propelled significant developments in a wide range of applications in science and engineering \cite{lecun2015}.
These include advancements in medical imaging \cite{greenspan2016guest, litjens2017survey}, computer vision \cite{he2016deep, dong2016image}, and artificial intelligence \cite{bengio2009learning}, with impressive performance in object recognition, natural language processing, and many other applications \cite{lecun2015, Bengio2013}.

%and advancements in face recognition \cite{schroff2015facenet, sun2014deep}, medical image analysis \cite{greenspan2016guest, litjens2017survey}, artificial intelligence \cite{bengio2009learning, arel2010deep} and many others.
Currently, most prominent applications of DL involve establishing complex decision boundaries in high dimensional parameter spaces using large amounts of training data.
We instead consider DL as a joint estimation framework for problems that contain unknown parameters in the measurement model.
Passive radar imaging falls into such class of problems, in which transmitter locations or transmitted waveforms may not be known a priori. %due to use of transmitters of opportunity, the measurement model is only known up to certain parameters such as transmitter location or transmitted waveforms.
%DL as a tool for joint estimation provides a paradigm to learn transmitter related parameters of the SAR imaging problem in order to reconstruct focused imagery.

Passive radar has been an area of intense research due to the proliferation of transmitters of opportunity and several advantages it offers. These include efficient use of electromagnetic spectrum, increased stealth, and reduced cost among others %by elimination of transmitter related hard-ware among many others
\cite{Yarman10, Malanowski09, Davidowicz12, Rapson12, Stinco13, Palmer13, Arroyo13, Malanowski14, wang2010passive, LWang12,Mason2015}. %\textbf{Please include other references from our research group. These are too limited. Include Doppler SAR, low-rank etc.}
Existing passive radar methods require two or more antennas, either spatially separated or colocated with sufficient directivity and gain.
Specifically, these methods rely on correlating pairs of measurements acquired by two different antennas.
These methods can be classified into two major categories: passive coherent localization (PCL) \cite{Baker05_1,Baker05_2,Wu2015,Hack2014_2,Kulpa12} and time-(or frequency) difference-of-arrival (TDOA/FDOA) backprojection \cite{Yarman10, Yarman08, Wang11, wang12, LWang12, Wang13_3, wacks2018doppler, Wacks14, Wacks14_2, Wang14, Qu2017}. %\textbf{Include our moving target and Doppler-SAR, Doppler-hitchhiker material as references.}

The PCL approach attempts to recover a copy of the transmitted waveform by filtering the received signal acquired by an antenna directed toward a transmitter of opportunity. This is followed by matched filtering of the received signal acquired by another antenna directed to a scene of interest \cite{Gogineni2016,Kulpa13,Kulpa2008,Feng2013}.
% in order to perform matched filtering with the back-scattered signal,
This approach relies on accurate estimation of transmitted waveforms. 
Recently, several algorithmic advances have been reported in waveform estimation using the structure of Digital Video Broadcasting-Terrestrial (DVB-T) signals as illuminators of opportunity %and requirement for high quality estimation of transmitted waveforms have motivated the recent developments in extracting noise-free reference signals using the waveform structure
 \cite{Hongchao2013,Baczyk2010,Xianrong2011,Mahfoudia2017, Baczyk2011}.
%Recently, single antenna PCL was proposed for DVB-T signals in \cite{mahfoudia2017feasibility}.  
%In \cite{Baczyk2011} an approach is developed to reconstruct the transmitted signal by demodulating the noisy signal to reconstruct the communication symbols.
%Their approach utilizes the structure of the guard band in DVB-T signals to determine the signal length.
%This is followed by a chirp-Z transform, used to estimate the modulation symbols, from which the signal can be reconstructed.
%An alternative approach is taken in \cite{Mahfoudia2017}, in which, the reference signal is extracted directly from the surveillance channel based on the signal structure, using the direct path component.
In addition to two antennas at each receiver location, and prior knowledge of the signal structure, PCL also requires direct line-of-sight to a transmitter of opportunity and high signal-to-noise ratio (SNR) for the received signal from the transmitter.

In the TDOA/FDOA backprojection approach, %use two or more sufficiently far apart receivers deployed on the same or different platforms.
received signals acquired by two or more sufficiently far-apart receivers are correlated and backprojected based on time or frequency difference of arrival to form an image of a scene \cite{Yarman08, Yarman10, Wang11, wang12, Wang13_3, wacks2018doppler, Wacks14, Wacks14_2}.
%Essentially, the phase distortion introduced by the unknown waveform is eliminated by this data correlation.
As compared to PCL, TDOA/FDOA backprojection does not require direct-line-of-sight to a transmitter, high SNR, or the knowledge of transmitter location.
However, the method is limited to imaging widely separated point scatterers.
To overcome this limitation, an alternative method based on low-rank matrix recovery (LRMR) has been developed \cite{Mason2015}.
Despite its effectiveness in reconstructing scenes with extended targets, the LRMR-based approach has significant computational and memory requirements, which preclude its applicability to realistically sized images.

Recently, the DL framework has been investigated for signal processing problems, specifically with an emphasis on sparse coding and compressed sensing.
In \cite{gregor2010}, the iterative soft thresholding (ISTA) and coordinate descent algorithms were implemented via a recurrent neural network (RNN), in which each layer of the network corresponded to an iteration.
%The model was trained in a supervised manner by \textcolor[rgb]{1.00,0.00,0.00}{target sparse codes} \textbf{what does this mean? Do not use expressions that are ambiguous in passive radar context} with the goal of accelerating convergence.
The model was trained in a supervised manner using the desired solutions for sparse codes of the corresponding inputs with the goal of accelerating convergence. 
This fundamental observation was exploited to implement an approximate message passing algorithm \cite{borgerding2016}, to learn problem specific gradient descent parameters \cite{andrychowicz2016learning}, and to estimate parametrized priors \cite{putzky2017recurrent}.
%Motivated by these advances, we investigated DL as a framework to combine model based imaging and data-driven approaches for synthetic aperture radar imaging \cite{mason2017deep}.  
In \cite{yonel2018deep} we extended the idea of emulating iterative maps using RNNs to image reconstruction problems in a Bayesian framework.
We considered the passive imaging problem in which the transmitter location is unknown, and used DL to refine the phase component of the synthetic aperture radar (SAR) forward model. %that result from transmitter location uncertainty in passive SAR.
Training was done in an unsupervised manner using SAR signals received directly via {complex backpropagation} \cite{leung1991complex}.
The method produces focused imagery without increasing the computational complexity of the proximal gradient descent algorithm which its based on.
\subsection{Our Approach and Its Advantages}

Following the principles introduced in \cite{mason2017deep, yonel2018deep}, we develop a DL-based approach for passive SAR image reconstruction when the transmitted waveforms of opportunity are unknown. 
The key advantage of our approach as compared to other methods is that it only requires a single receiver antenna, thereby providing reduced cost and increased versatility. 
Previously, we had presented preliminary results for joint waveform estimation and imaging for passive SAR in \cite{yonel2018deep2, yonel2018deep3}. 
In this paper, we present the deep network architecture and the unsupervised training scheme to learn transmitted waveforms as a parameter of the SAR imaging problem, while reconstructing focused imagery.
Specifically, we extend our preliminary studies by developing the theory of our approach and demonstrating its effectiveness via further numerical simulations. 

%We extend our preliminary studies on the joint waveform estimation and imaging problem in \cite{yonel2018deep2, yonel2018deep3}, develop the theory and demonstrate its effectiveness in numerical simulations.  

Unlike traditional PCL, our approach does not require an antenna directed to a transmitter of opportunity. 
We assume a stationary transmitter transmitting unknown waveforms and a single moving receiver collecting backscattered signal from a scene of interest. 
We then take an optimization perspective to DL, and interpret image reconstruction as a machine learning task.
We derive a proximal gradient descent update to solve for scene reflectivity, and formulate a recurrent-auto encoder that is parameterized by unknown waveform coefficients. 
As a result of our architecture, estimation of the transmitted waveforms is formulated as a parameter learning task via backpropagation. 
We use complex backpropagation to derive the parameter update equations, making our method applicable to both real and complex waveforms. 
Our method is based on unsupervised learning, in that, the model is trained solely on received back-scattered signals  without using any SAR images as labels. 
As a result, we avoid upper bounding the quality of reconstructed images by SAR images reconstructed using conventional methods.
The highly non-convex problem of backpropagation is guided to a feasible solution over the parameter space by initializing the network with the known components of the SAR forward model.

In our problem formulation, we assume that the structural form of transmitted waveforms are known \emph{a priori}, and represent them as a linear combination of known basis functions. 
A wide range of communication and broadcasting waveforms falls into such a class of waveforms including DVB-T and WiMAX signals.
We particularly formulate our method for transmitted signals with a flat spectrum, which is applicable to illuminators of opportunity generated from various spread spectrum methods. 
These include frequency phase shift keying (PSK) modulated, code division multiplexed (CDM) and orthogonal frequency division multiplexed (OFDM) signals.   %. and employ it as a constraint for waveform estimation. 
Recently in \cite{mahfoudia2017feasibility} a single receiver PCL methodology was proposed to estimate DVB-T signals for passive imaging, in which the signal structure was taken advantage of in processing the received signal from the surveillance channel.
However in our formulation, such prior knowledge on the waveform is used merely as a functional constraint during backpropagation, and lack thereof is not a limiting factor for the proposed framework.
Hence our approach provides flexibility to incorporate any prior information of the waveform structure to improve waveform estimation and imaging. 

%Moreover, prior information regarding the waveform structure is incorporated during network initialization and backpropagation.
% or a specific structure on the transmitted signal.
%We particularly formulate our method for transmitted signals with a flat spectrum, which is applicable to illuminators of opportunity generated from various spread spectrum methods. 
%These include frequency phase shift keying (PSK) modulated, code division multiplexed (CDM) and orthogonal frequency division multiplexed (OFDM) signals.   %. and employ it as a constraint for waveform estimation. 
%This formulation is applicable to illuminators of opportunity generated from a wide range of spread spectrum methods such as frequency phase shift keying (PSK) modulated, code division multiplexed (CDM) and orthogonal frequency division multiplexed (OFDM) signals.
%However, such prior knowledge on the waveform is used merely as a functional constraint in optimization, and lack thereof is not a limiting factor for the proposed framework.
%Instead, our approach provides a method to incorporate any prior information of the waveform structure to improve waveform estimation and imaging. 
%Waveforms used in communication and radar systems are designed with special purposes and thus have structure that is often known a priori.

Finally, in addition to the benefits of deploying a single receiver, our method provides means of estimating the waveform with the task of reconstructing enhanced imagery. %due to task-specificity of backpropagation learning. 
This is achieved by formulating waveform estimation as minimization of the mismatch between received SAR signal and the synthesized SAR signal from the reconstructed scene. %the SAR object of the reconstructed image in the range of the measurement map.
We show that our DL approach produces high contrast imagery when trained under realistic SNR levels. 
%We demonstrate that this is due to learning a sufficiently accurate flat spectrum waveform to the underlying QPSK signal. 
Our results indicate that the performance is strongly correlated to the accuracy of the estimated flat spectrum waveform, which demonstrates the joint estimation capability of the DL-approach. 

%\subsection{Organization of the Paper}

The rest of the paper is organized as follows. In Section \ref{sec2}, we introduce background material on deep learning. In Section \ref{sec3}, we present the received signal and waveform models. In Section \ref{sec4}, we present the network architecture for passive SAR image reconstruction. In Section \ref{sec5}, we discuss parameterization and training of the network.
We provide numerical simulations and discussion of the results in Section \ref{sec:simulation}.
Section \ref{sec:conclusion} concludes the paper.

We use lower case bold fonts to denote vector quantities in finite dimensional spaces and upper case bold fonts to denote matrices. 

%The paper denotes vector elements in lower case bold fonts, whereas upper case bold fonts denote matrices. Operators are denoted in upper case calligraphic symbols, and are parameterized by the surrogate symbol $\theta$ within brackets, and take inputs within square brackets. 
%The lower case bold $\x$ and italic bold $\bi x$ are reserved for the scene coordinates in discussion of the SAR forward model, and are specified in Section \ref{sec3}. 
%
%The rest of the paper is organized as follows. In Section \ref{sec2}, we introduce background material on deep learning. In Section \ref{sec3}, we formulate the data model for the waveform estimation problem. In Section \ref{sec4}, we present the network architecture for passive SAR image reconstruction. In Section \ref{sec5}, we discuss parameterization and training of the network. 
%We provide numerical simulations and discussion of the results in Section \ref{sec:simulation}. 
%Section \ref{sec:conclusion} concludes the paper. 

\section{Deep Learning Background}\label{sec2}

The most fundemental architecture in DL is the Artificial Neural Network, %(ANN),
%As previously stated, ANNs %depicted in Figure \ref{fig:network}, 
characterized by a cascade of affine mappings followed by point-wise nonlinear operations, referred to as \emph{layers}. 
Each layer produces a \emph{representation} $\h \in \Cb^M$ of its input $\mathbf{d} \in {C}^N$, defined as
\begin{equation}\label{eq:layerOutput}
  \h =  \sigma ( \mathbf{Ad + b} )
\end{equation}
where $\mathbf{A}\in \mathbb{C}^{M \times N}$ is the weight matrix, $\mathbf{b}\in \mathbb{C}^{M}$ is vector of corresponding biases, and $\sigma( \cdot )$ is an element-wise non-linear function, referred to as the activation function of the network.
%, and $\tilde{\f}$ is the representation produced by the layer. 
Letting $\Omega$ denote the input space, such that $\mathbf{d} \in \Omega$, the layer transforms $\Omega$ by \eqref{eq:layerOutput} to create a {feature space} containing the representation $\h$. 

%
%\begin{figure}
%\centering
%\includegraphics[scale=.3]{deep_net2.png}
%\caption{ A fully connected N-layer feed-forward neural network. Adapted from \cite{Goodfellow2016}. }
%\label{fig:network}
%\end{figure}

At each layer, a new representation of the previous layer output is generated, resulting in a hierarchical representation of the input. 
The output at the end of the $k^{th}$ layer can be written as
\begin{equation}\label{eq:singlelayeroutput}
\h_{k}  =  \sigma ( \mathbf{A}_{k}  \h_{k-1} + \b_{k} ).
\end{equation}
Letting $\phi : \Omega_L \rightarrow \Gamma $ be a mapping from the feature space produced by the $L^{th}$ layer, $\Omega_L$, to the output space $\Gamma$, and redefining $\d = [\d, \ 1]^T$ and $\W_k = [\mathbf{A}_k, \ \b_k]$, $k=0,...,L$, the network output $\mathbf{g}^* \in \Gamma$ becomes
\begin{equation} \label{eq:layeroutput}
\mathbf{g}^*  =  \phi\left(\W_L \sigma(\W_{L-1} ...  \sigma(\W_{1} \sigma(\W_{0} \mathbf{d})\big)\right).
\end{equation}
 
\eqref{eq:layeroutput} analytically defines the \emph{network operator}, $\mathcal{L}(\theta) : \Omega \rightarrow \Gamma$, which is the mapping between the input and output spaces, where
\begin{equation}\label{eq:NetworkOperator}
  \mathcal{L}(\theta) [\mathbf{d}] = \mathbf{g^*}, 
\quad  \theta = \{\mathbf{W}_k\}_{k = 1}^{L}.
\end{equation}

In summary, the weights of the network provide a parametrization of the operation that the network performs, whereas the non-linear unit introduces the capacity to approximate complex mappings between input and output spaces. 
The nested non-linear transformations are generally explained in terms of the universal approximation theorem or probabilistic inference \cite{Hornik1991,Cybenko1989, cooper1990computational}. 
The mapping performed by the network operator is referred to as \emph{forward propagation}. 

{Learning} procedure in the network is the estimation of $\theta$ with respect to a figure of merit given a set of training data $\left\{ \mathbf{d}_1, \mathbf{d}_2, \cdots,  \mathbf{d}_T \right\}$ and corresponding ground truth data set $G = \left\{ \mathbf{g}_1, \mathbf{g}_2 \cdots,  \mathbf{g}_T \right\}$. 
This is achieved by optimizing a cost function with respect to network parameters $\theta$, which typically defined as
%commonly chosen as the $\ell_2$-norm of mismatch at the network output:
\begin{equation}\label{eq:L2Error}
  \Jc_G [\theta] =  \frac{1}{2T} \sum\limits_{n=1}^T \|  \Lc(\theta)[\mathbf{d}_n] - \mathbf{g}_n \|^2_2.
\end{equation}
The analytic method of computing the derivatives through the network with respect to trainable parameters, $\theta$, is referred to as the \emph{backpropagation} algorithm. 
Network parameters $\theta$ are then updated via gradient descent such that
\begin{equation}\label{eq:updates}
\theta^{l+1} = \theta^{l} - \eta_l \nabla_\theta\Jc_G [\theta^l]
\end{equation}
where $\eta_l$ is the step size of the $l^{th}$ parameter update. 
 
For large training sets, the gradient term $\nabla_\theta\Jc_G [\theta^l]$ is estimated as an average of the gradient values computed over a small subset of the training data. 
This methodology, referred to as Stochastic Gradient Descent (SGD), performs several gradient updates each time the full data is used.
This update cycle is referred to as an "epoch."
% rather than a full gradient computation, and offers faster decay in the loss function from epoch to epoch than full gradient descent counter-part \cite{}. 

%the most prominent update method is the stochastic gradient descent(SGD). At each iteration, the update term is estimated by averaging the computed values of the gradient over a small subset of the training set. Going over the entire training set in computing updates completes an ``epoch''. 
%The optimization is most commonly carried out using stochastic gradient descent (SGD). 
%The analytic method of calculating the derivatives of the network with respect to trainable parameters, $\theta$, is referred to as the \emph{backpropagation} algorithm. 
The optimization over $\Jc_{G}[\theta]$ is typically a high-dimensional and non-convex problem.
The error surface often consists of many saddle points and local minima \cite{lecun2015,dauphin2014identifying}.
%This optimization problem is typically very high-dimensional and highly non-convex, often consisting of many saddle points and non-optimal local minima \cite{Goodfellow2016, lecun2015, dauphin2014identifying}.
As a result, a critical aspect of backpropagation is the initialization of the network parameters $\theta$, which is typically chosen to guide the network to a desirable locally optimal solution.  %by backpropagation. 

%Therefore, prior knowledge of the input to output mapping initializing the back propagation increases the chance of reaching an optimal solution.
%In this paper we propose to use physical forward model of SAR imaging to obtain good initialization.
We propose DL framework for problems in which we have unknowns or uncertainties in the measurement model. 
In general, we can model a measurement mapping as
\begin{equation}\label{eq:unknownModel}
\mathbf{d} = \mathbf{F}(\theta) \brho,
\end{equation}
where $\mathbf{d} \in \mathbb{C}^{M}$ are the measurements, $\brho \in \mathbb{C}^N$ is the unknown quantity to be recovered, and $\mathbf{F}(\theta)$ is the measurement map that depends on the parameters $\theta$. 
Following the DL framework for image reconstruction we introduced in \cite{yonel2018deep}, recovering the unknown $\brho$ can be interpreted as learning a representation of the measurements $\d$ in the image space.
In this sense, the DL framework captures the image reconstruction task at the forward propagation step. %in which representations formed by the network serve as an inversion tool. 
However, since $\theta$ is unknown and arbitrarily initialized, the reconstructed image is initially inaccurate.

\begin{figure}
\centering
\includegraphics[scale=1]{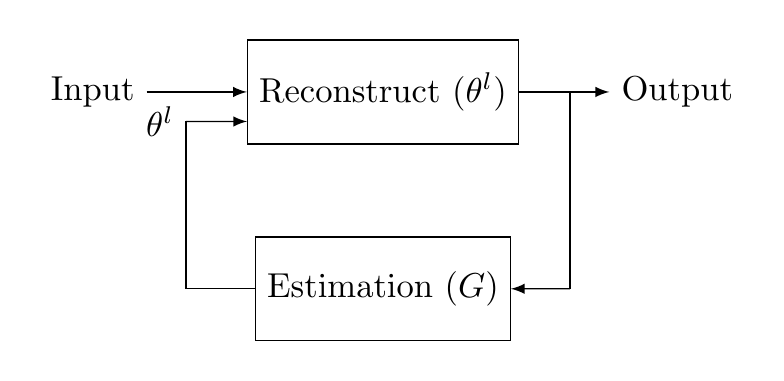}
\caption{\emph{The process flow of estimation via deep learning.} The forward propagation is modeled as the reconstruction method, whereas estimation is performed at backpropagation which acts as a feedback and updates network parameters.}
\label{fig:FigureBP}
\end{figure}

The advantages of the DL framework come at the backpropagation step, which allows the unknown parameters $\theta$ to be learned and to produce an accurate measurement map, thereby improving the accuracy of image reconstruction \cite{yonel2018deep}. 
A high-level illustration of the effect of the back-propagation step for imaging is provided in Figure \ref{fig:FigureBP}, in which the reconstruction step is parameterized by $\theta^l$, and estimation is parameterized by training data $G$ per \eqref{eq:L2Error}. 
Further details and discussion of DL framework for image reconstruction are provided in Sections \ref{sec4} and \ref{sec5}, while the effect of back-propagation in our specific application is illustrated in Figure \ref{fig:BpropFig}. 

%and hence an inversion. 
%that describe the precise measurement mapping in \eqref{eq:unknownModel}.
%As discussed in the following sections, we investigate the use of prior knowledge on $\theta$ and $\mathbf{F}$ in network initialization. %from the SAR measurement model in network initialization.
%With the initialization by the known components of the forward map, the network approximately becomes an inverse solver for the SAR problem.
%By training the inverse solver with SAR data sets, the forward model is refined.
%This in turn leads to an accurate inversion for image reconstruction. 
% and driven to become a more accurate solver for the inversion.
 
% by which we utilize backpropagation as a tool for refinement of the physical model. 

%\textcolor[rgb]{1.00,0.00,0.00}{Maybe include a couple paragraphs and maybe figures drawing a comparison with inverse problems}.

%It is interesting to understand that deep learning is inherently relate to inverse problems...

\section{Passive Synthetic Aperture Radar Imaging}\label{sec3}

\begin{figure*}[!ht]
\centering
\includegraphics[scale=.45]{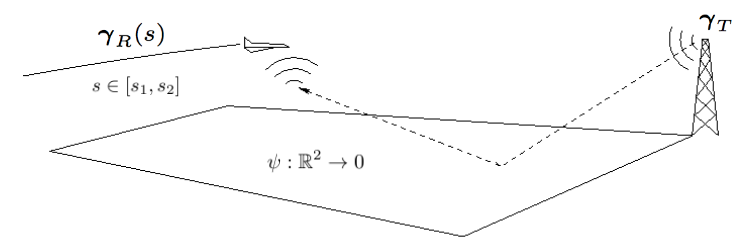}
\caption{A depiction of the passive SAR configuration with flat topography $\psi(\x)$. A stationary transmitter of opportunity located at $\bgamma_T$, $\bgamma_R : [s_1, s_2] \rightarrow \mathbb{R}^3$ denotes the receiver trajectory.}
\label{fig:sar_fig}
\end{figure*}

\subsection{Imaging Geometry}
Let $\x = [\bi x,\psi(\bi x)] \in \Rb^3$ denote the position of a scatterer, where $\bi  x = [x_1,x_2] \in \Rb^2$ and $\psi:\Rb^2 \rightarrow \Rb$ is a smooth function describing the ground topography. %Furthermore, $s \in [s_1,s_2]$ is the slow-time variable which parameterizes the location of the receiver. $\o\in[\o_1,\o_2]$ is the fast-time temporal frequency, and $c_0$ is the speed of light in free space.

We assume a stationary transmitter located at position $\bgamma_T \in \mathbb{R}^3$ and a moving receiver traversing a trajectory $\bgamma_R(s)$, where $s \in [s_1, s_2]$ denotes the slow-time variable.  
Figure \ref{fig:sar_fig} illustrates the passive SAR configuration under consideration. 
We assume that both the transmitter and receiver locations are known, but the transmitted waveforms are unknown. 
%In the measurement model we assume the transmitter location is known, but the transmitted waveform properties are not.

\subsection{Illimunators Of Opportunity}

This paper considers communication and broadcasting signals as the illuminators of opportunity. % are communication systems and use combinations of phase, frequency and amplitude modulation. 
These signals are designed based on spread spectrum methodology and characterized by orthogonal division of the time or frequency domain to transmit different communication symbols.
%The spread spectrum method that we focus our efforts towards are 
%We specifically consider orthogonal frequency division multiplexed (OFDM) waveforms.
One such class of signals are orthogonal frequency division multiplexed (OFDM) waveforms. 
OFDM signals involve different phase or amplitude modulated symbols that are transmitted using a set of orthogonal waveforms spanning the channel bandwidth \cite{Cimini1985,chen2009spectrum}.

OFDM signals %are commonly used in digital television and cellular communication, and 
have been widely studied as illuminators of opportunity due to the prevalence of DVB-T and WiMAX standards used throughout the world.  \cite{Rapson12,Zhao2013,Arroyo2011,Arroyo13,Gogineni2014,harms2010,Palmer13,Malanowski09,wang2010}.
%Studies show that the ambiguity function of the OFDM signal posses undesirable spikes resulting from the pilot tones of the signal.
Notably, the spectrum of an OFDM signal is relatively flat and noise-like over the channel \cite{harms2010, Palmer13}.
%This property is established from 
This characteristic can be observed by expressing the OFDM signal as a sum of random phase modulation symbols by applying the central limit theorem.
We assume that the transmitted waveforms may vary during the receiver's aperture time and model them as slow-time dependent as follows:
%For a more general representation, the transmitted waveform can be expressed in the form:
\begin{equation}\label{eq:wformRep}
W(\o,s) = \sum\limits_{k=1}^K c_k \varphi_k(\o,s),
\end{equation}
where $\varphi_k$, $k = 1, \cdots K$, are basis functions and $c_k$, $k = 1, \cdots K$, are (possibly complex) corresponding coefficients. %and $\varphi_k$ are the corresponding basis functions.
%Here we parameterize the waveform with the slow time variable to capture the scenario that the transmitted signal varies during data collection process. 
%Discretizing the fast time frequency $\o$ and slow time $s$ variables into $M$ number of pairs $\{ (\o, s)_m \}_{m = 1}^M$ and stacking the waveform elements $W((\o, s )_m)$ into a vector $\w \in \mathbb{C}^M$, we obtain the following representation:
%scheme as in obtaining \eqref{eq:datamodel3}, we can represent the waveform coefficient vector in \eqref{eq:datamodel3} as
%\begin{equation}\label{eq:general_waveform_model}
%\w = \boldsymbol{\Phi} \mathbf{c},
%\end{equation}
%where $\boldsymbol{\Phi}$ is a $M \times K$ matrix whose $k$-th column consist of the discretized basis function $\varphi_k((\o,s)_m)$, and $\mathbf{c}$ is a $K\times 1$ vector of coefficients $c_k$.
\eqref{eq:wformRep} provides a representation by which most common waveforms of opportunity can be modeled.
%Additionally, useful structure can be deduced from this representation to serve as constraints for estimation.
%For the case of phase modulated signals, the columns of $\boldsymbol{\Phi}$ are orthogonal and have unit norm, thus, the coefficients $c_k$ are unit-modulus, complex numbers. 

One such structure is that of binary or quadrature phase shift keying (B/QPSK) modulated signals which are building blocks of OFDM signals. 
Specifically, for QPSK, the basis coefficients are sampled uniformly on the unit circle, and the corresponding waveform has a flat spectrum.    %and $\varphi_k(\o, s)$ become standard basis functions. 
In our framework, we use this flat-spectrum structure as a statistical prior by the means of a constraint set in waveform estimation.
The same process can be used to model more complex waveforms to find constraint sets for estimation.

%In this case, the elements of $\w$ are each complex numbers with unit-modulus, and provide a constraint for estimation.
%The same process can be used to model more complex waveforms to find constraint sets for estimation with our framework.

\subsection{Passive SAR Forward Model}

Under the Born approximation and a flat topography assumption, we model the received signal as \cite{Yarman08_biSAR}
\begin{equation}\label{eq:datmodel2}
d(\o, s) \approx  W(\o, s) \int \mathrm{e}^{-\mathrm{i}\frac{\o}{c_0}R(s,\bi x)} a(\bi x, s) \rho(\bi x) d\bi x, 
\end{equation}
%Moving the unknown waveform component out of the $\bi x$ integral, 
%$$
% W(\o, s) \mathcal{\tilde{F}}(\o,s) [\rho] $$
%where $\mathcal{\tilde{{F}}}(\o, s):\mathbb{R}^2 \rightarrow \mathbb{C}$ corresponds to the known component of the forward model, 
%  \begin{equation}
%d(\o,s) \approx \Fc[\rho](\o,s) := \int \mathrm{e}^{-\mathrm{i}\frac{\o}{c_0}R(s,\bi x)} A(\o,s,\bi x) \rho(\bi x) d\bi x
%\label{eq:FIO_forward_model}
%\end{equation}
%$R(s,\bi x)$ is the bistatic range given by
%W(\o,s) 
%\begin{equation}
%\phi(\o,s,\bi x) = \frac{\o}{c_0}R(s,\bi x)
%\end{equation}
%where $c_0$ is the speed of light in free space and $R(s,\bi x)$ is the range travelled by electromagnetic waves.
%For mono-static SAR, it is given as
where
\begin{equation}
R(s,\bi x) = |\bgamma_T - \x| + |\bgamma_R(s)-\x|
\label{eq:mono_range}
\end{equation}
is the bistatic range, $\o \in [w_1, w_2]$ is the fast-time frequency, $c_0$ is the speed of light in free-space, $W(\o, s)$ is the waveform transmitted at $s \in [s_1, s_2]$, and $a(\bi x, s)$ is the azimuth beam pattern. %models the receive antenna and azimuth beam patterns, and geometric spreading factors.
We let
\begin{equation}\label{eq:knownComp}
\mathcal{\tilde{F}}(\o,s) [\rho]  := \int \mathrm{e}^{-\mathrm{i}\frac{\o}{c_0}R(s,\bi x)} a(\bi x, s) \rho(\bi x) d\bi x
\end{equation}
and write \eqref{eq:datmodel2} as
%Having moved the unknown waveform component out of the $\bi x$ integral in equation \eqref{eq:datmodel2}, we can represent the measurements:
\begin{equation}
d(\o, s) \approx W(\o, s) \mathcal{\tilde{F}}(\o,s) [\rho].
\end{equation}
%where $\mathcal{\tilde{{F}}}(\o, s):\mathbb{R}^2 \rightarrow \mathbb{C}$ corresponds to the known component of the forward model as:
%In most applications the antennas are sufficiently broadband, meaning that the amplitude is approximately constant over illuminated region.
Without loss of generality, we assume $a(\bi x, s)$ is constant and set it to $1$. 
%Thus, in a spot-light configuration for high resolution SAR imaging, the same region stays in the main-lobe of the antennas and the amplitude remains approximately constant over the aperture. 
%Under these considerations the antenna beam patterns and geometric spreading are approximately constant and we set $a(\bi x, s) \approx C_{RT}$, where $C_{RT} > 0$ is a constant, typically of value $1$.
Placing the origin at the center of the scene and Invoking the far-field and small scene approximations for a sufficiently long aperture, $\mathcal{\tilde{F}}$ can be further approximated to yield the following measurement model: %of the spotlight-SAR geometry on the phase component of the forward model, one obtains the plane wave-front approximation on the phase \eqref{eq:mono_range} of operator $\tilde{F}$, such that:
\begin{equation}\label{eq:SpotSAR}
d(\o, s) \approx W(\o, s) \mathrm{e}^{- \mathrm{i} \frac{\o}{c_0}( | \bgamma_R(s)| + | \bgamma_T |)} \ \int \mathrm{e}^{-\mathrm{i}\frac{\o}{c_0} \mathbf{k} \cdot \bi x} \rho(\bi x) d\bi x,
\end{equation}
where $\mathbf{k} = \widehat{\bgamma_R(s)} + \widehat{\bgamma_T}$ \footnote{$\hat{\x}$ denotes the unit vector in the direction of $\x$}, and $\mathcal{\tilde{F}}$ is now approximated by the Fourier transform up to a phase factor.  

%where $\bgamma_R(s)$ and $\bgamma_T$ denote the receiver and transmitter locations, respectively. 
%In this paper, we consider a problem setting where the transmitter location is stationary and known, while the transmitted waveforms and antenna beam patterns are unknown. 
%Furthermore, we assume that the waveform may vary with respect to slow time variable $s$, and model it as $W(\o,s)$. 
%Factoring out the waveform from \eqref{eq:forward_amp}, we have $A(\o,s,\bi x) = W(\o, s) \tilde{A}(\o, s,\bi x)$. 

We discretize the scene into an $N$-pixel grid of $X =\{\bi x_i\}_{i=1}^N$, and stack $\rho(\bi x_i)$, $i = 1, \cdots N$ into a vector $\brho \in \mathbb{C}^N$.
%%\begin{equation}\label{eq:datmodel2}
%%d(\o, s) \approx W(\o, s) \tilde{{F}}(\o,s) [\brho]
%%\end{equation}
%where $\tilde{{F}}(\o, s):\mathbb{R}^2 \rightarrow \mathbb{C}$ corresponds to the integral component of \eqref{eq:FIO_forward_model}.
% under the assumptions that we have a known transmitter location and a broadband antenna at the receiver. 
%The data and the waveform are discretized by sampling $\o$ and $s$ into $\{(\o, s)_j \}_{j=1}^{M}$ pairs, and stacked into vectors $\d, \w \in \mathbb{C}^M$.
We discretize the fast-time frequency $\o$ and slow-time $s$ variables into $\{(\o, s)_m \}_{m=1}^{M}$ pairs, and stack SAR measurements into a vector $\d \in \mathbb{C}^M$.  
%The problem is represented in finite dimensional space after discretizing the data and the waveform by sampling $\o$ and $s$ into $\{(\o, s)_i \}_{i=1}^{M}$ pairs. 
%The data samples correspond to $(\o,s)_j$ pairs, which are stacked into a vector $\d \in \mathbb{C}^M$.
With the same sampling scheme, we stack the sampled waveform elements $W(\o, s)\vert_{(\o,s)_m}$ into a vector $\w \in \mathbb{C}^M.$%the unknown waveform is represented using the standard basis, and the unknown quantity are the basis coefficients stacked in vector $\w \in \mathbb{C}^M$, by which equation \eqref{eq:datmodel2} becomes:
%With this representation the unknown waveform is equal to the coefficients under the standard basis stacked in vector $\w \in \mathbb{C}^M$, by which equation \eqref{eq:datmodel2} becomes:
Thus, we obtain
\begin{equation}\label{eq:datamodel3}
\d \approx \text{diag}(\w) \tilde{\mathbf{F}} \brho, 
\end{equation}
%In \eqref{eq:datamodel3}, $\w \in \mathbb{C}^M$ is the discretized and vectorized waveform such that $\w_m = W(\o, s)\vert_{(\o,s)_m}$, and 
where $\tilde{\mathbf{F}}$ is the matrix corresponding to the finite dimensional representation of $\mathcal{\tilde{F}}$ in (\ref{eq:knownComp}). 
%Note that the transmitter is stationary, and its location $\bgamma_T$ is may be unknown, neither is the transmitted waveforms nor the transmitted antenna beam patterns. As a result the forward model $\mathcal{F}$ is only partially known.
%Under a fully known forward model $\F$, a bistatic SAR image with good geometric ﬁdelity can be formed by a two-layer filtered-backprojection type operation \cite{Yarman08}.

%\subsection{Waveform Modeling}
%
%In the previous subsection we assumed the waveform was represented in the standard basis. 
%However, typical radar and communication waveforms possess structures chosen to maximize information transmitted or optimize certain radar tasks that can better be represented in a parametrized form. 
%%For example, in communication systems different modulation schemes are used, ranging from phase modulations to spread spectrum methods.
%%In radar, waveforms are typically chirped to maximize resolution or designed to minii
%Thus, if we represent the waveform as
%\begin{equation}
%W(\o,s) = \sum\limits_{m=1}^M c_m \varphi_m(\o,s)
%\end{equation}
%where $c_m$ serve as the coefficients and $\varphi_m$ serve as basis functions, discretizing the fast-time and slow-time variables.
%
%In the case of modulation, such as phase shift keying, we set $c_m = e^{N}$ and $\varphi = \delta(\o-\o_m)\delta(s-s_m)$.

%\subsection{Waveform Structure}

\subsection{Image Formation}

Without complete knowledge of the forward model matrix $\F = \text{diag}(\w) \tilde{\F}$, we are no longer able to form bistatic SAR images using a two-layer filtered-backprojection type operation such as the one described in \cite{Yarman08_biSAR}.
%Similarly the unknown components in $\F$ compromise the effect of optimization based methods, since each iteration would consist of projections by an inaccurate operator and the error would be compounded. 
%Thus, we propose the use of a machine learning method to handle the uncertainties of the forward model, specifically, a deep learning network.
Similarly, optimization-based reconstruction approaches are not applicable due to unknowns in the forward model. 
Since the dependence of the waveform coefficients in the forward model is multiplicative in the frequency domain, the problem of image reconstruction can be viewed as a {blind deconvolution} problem.
A popular approach to solve such problems is to use an alternating minimization scheme, which requires solution of two minimization problems at each iteration. 
In this work, we instead propose a data driven approach based on DL. % based on DL. %, described in the following sections. 
The main advantage of the DL-based method is that it is task-driven. 
Instead of casting the unknown waveform as a joint parameter of the objective function in optimization, we cast it as a parameter of the optimizer. 
%This allows for estimating a waveform in a manner that will specifically produce enhanced imagery.  
This results in conducting waveform estimation to specifically produce accurate imagery. 

If the waveforms were known, estimation of the scene reflectivity could be formulated within a Bayesian framework as the following optimization problem:
%Thus, we formulate the problem as minimization of
%\begin{equation}\label{eq:objFunc_forward}
%\brho^* = \underset{\brho}{\text{argmin}} \ f(\brho) =  f_1(\brho) + f_2(\brho).
%\end{equation}
%The objective function $f(\brho)$ can be derived within Bayesian framework, in which case, $f_1$ models the log-likelihood and $f_2$ is the log-prior capturing our prior knowledge of the unknown.
%\textbf{Bariscan, in the previous lines you use function, in the next line you use functionals. Pick one and be consistent. I am inclined to say function as we are in finite dimensions, however, the calligraphic letters with square brackets imply functional, and the regular $f$ with square brackets seems odd. I will not change these around, you will need to address this.}
%In general, the objective $\mathcal{J}[\brho]$ has the form of the sum of two functionals such that
%\begin{equation}\label{eq: ObjFuncForm}
% \mathcal{J}[\brho] := f_1[\brho] + f_2[\brho],
%\end{equation}
%Under Gaussian noise assumption, we write the solution to \eqref{eq:objFunc_forward} as:
\begin{equation}\label{eq:obj_form_prob}
 \brho^* =\underset{\brho}{\text{argmin}} \ \frac{1}{2} \| \d - \text{diag}(\w) \tilde{\F} \brho \|_2^2 + \lambda \Phi(\brho), 
\end{equation} 
%\textbf{Bariscan, you have $f_2$ then introduce $\Phi$, then go back to $f_2$, consider if this is necessary and make notation consistent. }
where the $\ell_2$-norm term represents the log-likelihood function under an additive white Gaussian noise assumption, $\Phi(\brho)$ is the regularizer capturing the prior information on $\brho$, %, which results from the probabilistic prior defined on the unknown, 
and $\lambda>0$ is the regularization parameter. 

By deploying a convex $\Phi$, the optimization can be implemented as a forward-backward splitting algorithm \cite{combettes2005signal}.
%In this general formulation, the regularizer can be chosen to as either a smooth or non-smooth function.
%\emph{proximal gradient descent} \cite{} methods by applying \emph{forward backward splitting}. 
%These methods pursue optimization of the objective function in (\ref{}) by splitting the components $f_1$ and $f_2$, and derive the algorithm by operations resulting individually from each component \cite{}. 
The forward-backward splitting algorithm takes the form of a gradient descent step over the smooth $\ell_2$-norm term, followed by a projection onto the feasible set defined by $\lambda \Phi$ as follows: 
%taking an $\alpha$ sized step in the negative gradient direction over the smooth $f_1$ term, followed by a mapping by the \emph{proximity operator} of the non-smooth $f_2$ term to yield the iteration:
\begin{equation}\label{eq:proxgdes}
\brho^{k+1} = \mathcal{P}_{\alpha \lambda \Phi} ( \brho^k - \alpha \nabla f[\brho^k] ), 
\end{equation}
where
\begin{equation}\label{eq:objFunc_f}
f(\brho) : =  \frac{1}{2} \| \d - \text{diag}(\w) \tilde{\F} \brho \|_2^2,
\end{equation}
$\alpha >0$ is the step size, and $\mathcal{P}_{\alpha \lambda \Phi}(\cdot)$ is the \emph{proximity operator} of $\alpha \lambda \Phi(\brho)$ defined as
\begin{equation}\label{eq:proxop}
\mathcal{P}_{\alpha \lambda \Phi}(\brho) = \underset{\y}{\text{argmin}} \ \frac{1}{2} \| \y - \brho \|_2^2 + \alpha \lambda \Phi (\y).
\end{equation}
When $\Phi$ is the indicator function of a closed convex set, the proximity operator is simply an orthogonal projection onto that set.
%While the analysis of forward-backward splitting algorithm only applies to convex $\Phi$, there are classes of non-convex regularizers that produce iterations similar to \eqref{eq:proxgdes}.
%The most common example involves the $\ell_0$-norm as the regularizer which leads to the iterative hard thresholding algorithm \cite{blumensath2007iterative}.  %which uses the $\ell_0$ norm as the regularizer. 
%Although, that iteration follows from a more general algorithm that proves convergence results similar to forward-backward splitting for penalty functions which are semi-algebraic.
Using \eqref{eq:objFunc_f} and inserting $\nabla f[\brho^k]$ into \eqref{eq:proxgdes}, we have % using the objective function in \eqref{eq:obj_form_prob}, we have:
\begin{equation}\label{eq:ourUpdate}
\brho^{k+1} = \mathcal{P}_{\alpha \lambda \Phi} ( (\mathbf{I} - \alpha \tilde{\mathbf{F}}^H \text{diag}(| \w |^2) \tilde{\mathbf{F}}) \brho^k 
+ \alpha \tilde{\mathbf{F}}^H \text{diag}(\w)^H \d ), 
\end{equation} 
where $| \w |^2$ denotes element-wise magnitude square of $\w$. 
%Of course, \eqref{eq:ourUpdate} cannot be accurately performed since $\w$ is unknown, however this serves as the foundation for development of our DL approach, as discussed in Sections \ref{sec4} and \ref{sec5}. 
Clearly, iterative reconstruction in \eqref{eq:ourUpdate} cannot be used without the knowledge of $\w$. %which is unknown. 
Nevertheless, \eqref{eq:ourUpdate} serves as the foundation for the development of our DL-based approach discussed in Sections \ref{sec4} and \ref{sec5}. 

\section{Network Architecture}\label{sec4}
Our goal is to implement a deep network that can recover scene reflectivity accurately when the waveform coefficients are unknown. 
To achieve this, we deploy the methodology of \cite{yonel2018deep} for SAR imaging to simultaneously estimate the waveform and reconstruct the image. 
%The motivation of this architectural choice is simple yet significant. 
Specifically, we use a recurrent auto-encoder architecture \cite{rolfe2013}. 
A recurrent auto-encoder consists of a two stage network: a recurrent neural network (RNN) that emulates the iterative imaging algorithm defined by \eqref{eq:ourUpdate}, and a \emph{decoder} stage that maps the image estimate back to the measurement space. 
The main advantage of this architecture is that it allows for unsupervised training using the SAR received signal.

\subsection{RNN-Encoder}

RNNs fundamentally differ from other architectures since the parameters are shared over layers. 
As a result, in an RNN, each layer performs the same transformation between feature spaces. 
The RNN encoder is designed by {unfolding} the update equation in \eqref{eq:ourUpdate} for a fixed number of iterations, say $L$. 
This emulates the imaging algorithm described in Section \ref{sec3}.
%such that the network non-linearity is the proximity operator\footnote{Note that the regularizer must have closed form proximity operator that acts element-wise to construct a neural network.} that results from the regularizer in \eqref{eq:obj_form_prob}.
The corresponding weight matrix and the bias vector are defined as
\begin{equation}\label{eq:networkArch}
\mathbf{Q} = \mathbf{I} - \alpha \tilde{\mathbf{F}}^H \text{diag}(| \w |^2) \tilde{\mathbf{F}}, \quad \b = \alpha \tilde{\mathbf{F}}^H \text{diag}(\w)^H \d,
\end{equation}
%and
%\begin{equation}
%\b = \alpha \tilde{\mathbf{F}}^H \text{diag}(\w)^H \d,
%\end{equation}
respectively. 
With this approach, the iterations become the feature transformations performed by the layers in \eqref{eq:layerOutput} and we get the following update equation:
\begin{equation}\label{eq:network_layer}
\brho^{k+1} = \P_{\alpha \lambda \Phi}(\mathbf{Q}\brho^k + \alpha \mathbf{F}^{H}\mathbf{d}), 
\end{equation}
where $\text{diag} ( \w ) \tilde{\mathbf{F}} = \mathbf{F}$.
Thus, the image estimates $\brho^k$, $k = 1, \cdots L$, become representations produced at the $k^{th}$ layer of the network and the proximity operator $\mathcal{P}_{\alpha \lambda \Phi}(\cdot)$ becomes the network activation function. 
The only condition required on the regularizer to construct a neural network is that it must have a closed-form proximity operator that acts element-wise on its argument. 

\begin{figure*}[!ht]
\label{fig:recurrentautoenc}
\centering
\includegraphics[scale=0.8]{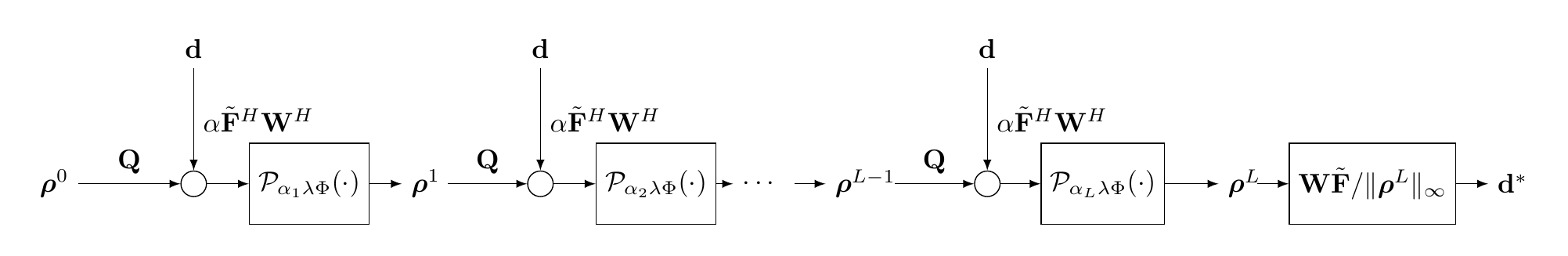}
\caption{Proposed recurrent auto-encoder architecture with $\mathbf{W} = \text{diag}(\w)$. The initial estimate $\brho^0$ can be set according to preference(set as vector of zeros in this work). Following the estimate generated at layer $L$, a linear decoder maps the estimate back to the data space. Figure depicts the case in which different learning rates ($\alpha_i$'s) are used at each iteration of the forward solver.}
\end{figure*}

%We initialize the network weights and biases with the known components $\tilde{\mathbf{F}}$.
%Furthermore, an a priori functional form for the initial waveform coefficients $\w^0$, and properly chosen step-size $\alpha$, the network becomes a solver for the convex minimization problem in \eqref{eq:obj_form_prob} at forward propagation, generating the estimate $\brho^*$ after $L$ layers. 

While we motivate the network architecture from an optimization perspective, we modify the forward propagation in a way that deviates from this analogy. 
Notably, optimization by \eqref{eq:network_layer} is conducted over the complex domain to estimate the scene reflectivity. 
Furthermore, the proximal gradient method that our optimizer is based on requires a large number of iterations to converge. %for convergence. 
However, such a large number of layers is not practical from the point of view of training. 
Hence, we set $L$ much smaller than the number of iterations needed for convergence. 
Additionally, we require the output of each layer to form visual representations of the received SAR signal. 
This is achieved by removing the phase of the representations in the network at forward propagation. 

In the scope of this paper, we consider the recovery of sparse scenes and use the $\ell_1$-norm constraint for $\Phi$ as the sparsity inducing prior.
%The use of $\ell_1$-norm constraint is not a limitation of our framework, as formulation with any $\Phi$ that has a well-defined, element-wise proximity operator leads to the same iterative form in \eqref{eq:network_layer}. 
The use of $\ell_1$-norm constraint is not a limitation of our framework, as the same iterative form in \eqref{eq:network_layer} is obtained with any $\Phi$ that has a well-defined, element-wise proximity operator. 
Furthermore, if the underlying scene is not sparse, a sparsifying transform such as the wavelet transform can be deployed, and optimization can be re-formulated as recovery of sparse coefficients. 
%The weight and bias terms in \eqref{eq:networkArch} would then be represented by the composition of the forward map and the sparsifying transformation matrix.
%The optimization can then be re-formulated as recovery of sparse coefficients, in which weight and bias terms in \eqref{eq:networkArch} are represented by the composition of the forward map and the sparsifying transformation matrix.
Therefore, we define the network non-linearity as a {phaseless} {soft thesholding operator} as follows:
\setlength\belowdisplayskip{12pt}
\begin{equation}
\mathcal{P}_{\tau \ell_1}(\brho)  =  \text{max} (|\brho| - \tau, 0 ) , \quad \brho \in \mathbb{C}^N, 
\end{equation}
where $| \brho |$ denotes taking the absolute value of the entries of $\brho$, $\tau$ is the threshold determined by scaling the $\ell_1$-norm constraint. 
With this modification, every representation in the RNN becomes a visual image, and feature mappings in \eqref{eq:network_layer} can be interpreted from an image processing perspective, as discussed in Section \ref{sec5}. 

\subsection{Decoder}
%
%In implementing the decoder, we consider that the scene reflectivity may be upper bounded given the operating frequencies of the receiver and typical scene refractive indices.
%Without loss of generality, we assume that the scene reflectivity varies between 0 and 1 and normalize the final RNN output ${\brho}^L$, before projection onto the data space as follows:
In implementing the decoder, we consider that scene reflectivities may be upper bounded given the operating frequencies of the receiver and typical scene refractive indices.
Without loss of generality, we assume that the scene reflectivities vary between 0 and 1, and that the scene consists of at least one strong reflector.
Under these assumptions, we normalize the final RNN output ${\brho}^L$ before projection onto the measurement space as follows:
\begin{equation}\label{eq:NormalizationFP}
{\brho}^* = \frac{{\brho}^L}{\ \ \| {\brho}^L \|_{\infty}}.
\end{equation}
The normalization of the final output enhances the effect of learning in light of the expected range of reflectivity values in the reconstructed image.

Following the normalized image formation step, the decoder maps the estimate $\brho^*$ back to the received signal space and synthesizes an estimate of the {SAR measurement} as follows: %The measurement synthesis corresponding to the image estimate $\brho^*$ is its representation on the range of the SAR forward model such that
\begin{equation}\label{eq:linlayer}
\mathbf{d}^* = \text{diag}(\w) \tilde{\mathbf{F}} {\brho}^*.
\end{equation}
 
By the insertion of the linear decoding layer \eqref{eq:linlayer}, the network operator $\mathcal{L}[{\theta}]$ acts as an approximation to the identity map on the received signal space. Representing $\brho^*$ as a function of network parameters $\theta$ and the input $\mathbf{d}$, we write
\begin{equation}\label{eq:networkParm}
\mathcal{L}[\theta](\d) = {\d}^* :=  \text{diag}(\w) \tilde{\mathbf{F}} {\brho}^*(\theta, \d).
\end{equation}
%where $\theta$ denotes the network parameters. % $\theta = \{ \mathbf{Q}, \mathbf{b} \}$.

%\textcolor[rgb]{1.00,0.00,0.00}{\textbf{Bariscan, Not sure if the level of detail in the items below is necessary. Or this could be summarized as a few sentences following the paragraph. Specifically, I think point 2 either belongs in the next section, or is too much info for a conference paper. Stuff to consider...}}

The final mapping back to received signal space allows the model to be trained in an \emph{unsupervised} manner. 
%To estimate the parameters $\theta$ that will improve reconstruction performance, we only need a training set of SAR measurements. 
%This formulation introduces two critical properties to our framework: 
This formulation offers a significant advantage. % for our framework. 
A supervised training scheme would require ground truth images coupled with SAR measurements. 
However, in the context of image reconstruction, a large number of SAR images acquired using the same imaging geometry may not be available. 
Moreover, training the RNN using %that performs the data to image space mapping 
SAR images would upper bound the performance by the reconstruction quality of conventional imaging algorithms that formed the SAR images in the first place.
Our approach avoids these drawbacks by unsupervised training. 

\section{Network Training}\label{sec5}
%In discussing the training, we analyze the three significant components of the procedure: forward propagation, initialization, and backpropagation. We first contemplate how SAR imaging is performed with the network in forward propagation, in the hypothetical case that the measurement mapping in \eqref{eq:datamodel3} is fully known with the underlying waveform coefficients. We then discuss how random initialization of the waveform coefficients diminish the functionality of the network operator, with specific emphasis on target localization and resolution. We then finally lay out the parameterization of the network and present the means of estimating the underlying waveform by unsupervised training of the model.

\begin{figure}
\centering
\includegraphics[scale=0.82]{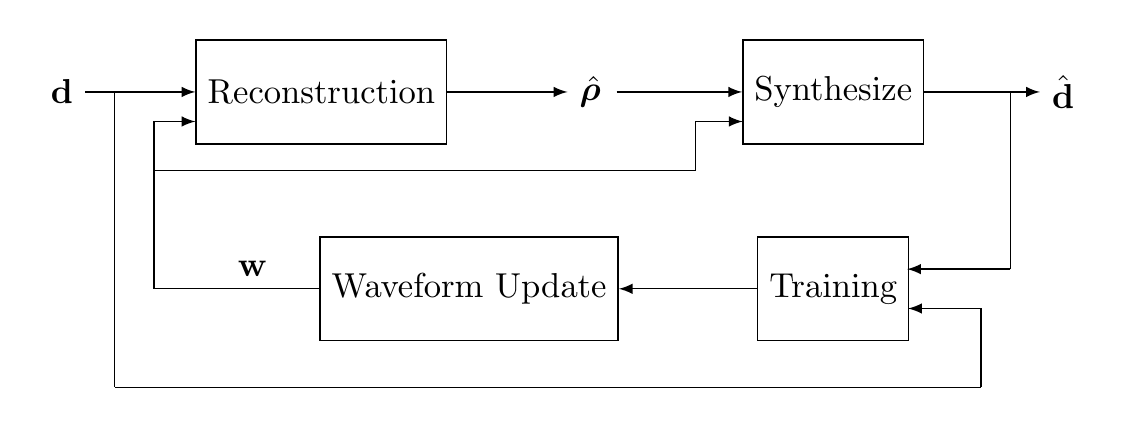}
\caption{\emph{The process flow of the proposed recurrent auto-encoder.} The optimizer reconstructs images, which are used to synthesize the input received signal. The mismatch is back-propagated to update the shared parameters of the two modules.}
\label{fig:BpropFig}
\end{figure}

\subsection{Forward Propagation For SAR Imaging}\label{sec:ForwProp}

Consider the ideal expressions for the network weight matrix and bias terms in \eqref{eq:networkArch} with the true underlying waveform.
%The normal operator $\mathbf{F}^H \mathbf{F}$ is a spatially-varying filter of low pass characteristic.
Essentially the map $\mathbf{Q}$ is composed of an all pass filter $\mathbf{I}$ of scale one, and $\mathbf{F}^H \mathbf{F}$, which is a spatially varying filter of low pass characteristic. %low pass filter of scale $-\alpha M$.
Having $\alpha M \ll 1$, $\mathbf{Q}$ acts as an image domain filter that gradually suppresses all frequencies by a $1-\alpha M$ factor, except for the high frequency bands determined by the cutoffs of the rows of $\mathbf{F}^H \mathbf{F}$. 
By the definition of $\mathbf{b}$ in \eqref{eq:networkArch}, the result of the linear filtering operation is \emph{biased} by the backprojection estimate. % by addition of the term $\b = \alpha \mathbf{F}^H \d$. 
Since $\tilde{\mathbf{F}}$ is sampled from a Fourier Integral Operator \cite{Yarman08_biSAR}, we know that backprojection preserves the target placement and edges in the image.
Hence, the biasing step practically enhances the foreground of the output of the filtering operation. 
By repetitive application of $\mathbf{Q}$ and biasing, mid to low frequency components are gradually suppressed, whereas the edges get further enhanced due to their high frequency content. 
%As the mid to low frequency components are gradually suppressed due to repetitive application of $\mathbf{Q}$ through the network, the edges get further enhanced by the biasing due to their high frequency content. 
Thereby, pixel-wise absolute-value thresholding at each layer effectively performs background suppression rather than suppressing the foreground, and the composite map of the layers becomes a non-linear enhancement filter. 
  
However, as emphasized throughout the paper, the forward map $\mathbf{F}$ is not fully known and at initialization the network cannot perform these operations accurately due to an arbitrary initial $\w$. 
%Moreover, the key property of backprojection from Fourier Integral Operator theory is significantly diminished with a randomly initialized waveform. 
Let $\w = {{\w^0}}$ be a randomly picked waveform coefficient vector to initialize the network parameters $\mathbf{Q}$ and $\b$. 
Setting the initial image estimate fed to the network as $\brho^0 = 0$, the representation at the first layer becomes
\begin{equation}
\brho^1 = \text{max} ( | \alpha \tilde{\mathbf{F}}^H \text{diag}({\w^0})^H \d | - \alpha \lambda , 0 ).
\end{equation}
%At the intermediate step resulting from the backprojection, we define $\tilde{\brho}$ as
We define $\tilde{\brho}$ as the intermediate estimate resulting from the backprojection with its $j^{th}$ pixel value given as
\begin{equation}\label{eq:backprojpix}
\tilde{\brho}_j := [\tilde{\mathbf{F}}^H \text{diag}({\w^0})^H \d]_j = \sum_{k = 1}^M (\overline{{\w}}^0)_k \overline{\left(\tilde{\mathbf{F}}\right)}_{kj}  \d_k , \ \ j = 1, \cdots, N.
\end{equation}
Plugging in \eqref{eq:datamodel3} into \eqref{eq:backprojpix}, and breaking down the contributions of pixels with indices $i=j$ and $i \neq j$ into the estimate of the reflectivity at the $j^{th}$ pixel, we obtain
%\begin{equation}
%\tilde{\brho}_j = \sum_{k = 1}^M \sum_{i = 1}^N   \overline{\left(\tilde{\mathbf{F}}\right)}_{kj} \left(\tilde{\mathbf{F}}\right)_{ki} (\overline{{\w}}^0)_k ({\w_o})_k  \ \brho_i
%\end{equation}
%The estimate of the reflectivity at pixel $j$ can be broken down into contributions of $i = j$, and $i \neq j$, such that:
\begin{equation}\label{eq:wformcorr}
\tilde{\brho}_j  =  \brho_j \left( \sum_{k = 1}^M  (\overline{{\w}}^0)_k ({\w_t})_k \right) +  \sum_{k = 1}^M \sum_{i \neq j} \overline{\left(\tilde{\mathbf{F}}\right)}_{kj} \left(\tilde{\mathbf{F}}\right)_{ki} (\overline{{\w}}^0)_k ({\w_t})_k  \ \brho_i
\end{equation}
%approximating $|A(s, \omega, \x)|^2$ as a constant amplitude such that $C_{RT} = 1$ in $\tilde{\mathbf{F}}$ in \eqref{eq:datmodel2}. 
where $\w_t$ is the vector of true waveform coefficients, and $\brho_j$, $j = 1, \cdots N$ are the elements of the  true scene $\brho$. 

%The immediate implication of an inaccurate initialization 
%The significance of the initialization is that the contribution of the ground truth pixel $\brho_j$  in its estimate $\tilde{\brho}_j$ is scaled by the correlation between the initial guess and the true waveform coefficients.
Consider a passive imaging scenario such that the true waveform coefficients are sampled from a QPSK signal, and that we set entries of $\w^0$ with randomly picked symbols from the unit circle in $\mathbb{C}$. 
Since such a randomly initialized $\w^0$ is highly unlikely to be correlated to the true QPSK waveform,
%The key impact of such inaccurate initialization is that the low correlation between $\w_t$ and $\w^0$ diminishes 
the contribution of the underlying scene pixel $\brho_j$ to its estimate $\tilde{\brho}_j$ diminishes.
Furthermore, this contribution is diminished to a level comparable to that of other points in the underlying scene $\brho_i$ where $i \neq j$.
As all scene reflectivities are scaled by low correlated complex exponentials in \eqref{eq:wformcorr}, targets get suppressed and a noisy image is obtained by backprojection. 
This phenomenon is observed in Figure \ref{fig:Figure1} from the reconstruction performance with a randomly initialized waveform. % as reconstruction with randomly initialized waveform fully suppresses the target and yields an image representative of noise. 
Therefore, obtaining a highly correlated waveform coefficient vector $\w$ is the key component of the reconstruction process by \eqref{eq:network_layer}. 

\subsection{Backpropagation For Waveform Estimation}
%The significance of the approach arises in the backpropagation step.
% which is achieved by parameterization of the network.  
Instead of parameterizing the feature maps of the RNN by the weight matrix $\mathbf{Q}$ and the bias vector $\b$, we limit parameterization to only the waveform vector $\w$. 
Thereby, backpropagation is cast as a solver for the waveform estimation problem. 
Furthermore, splitting the waveform as a parameter preserves the known components of $\mathbf{Q}$ and $\b$, which are initialized using $\tilde{\mathbf{F}}$. % from \eqref{eq:datmodel3}.   

%Conventionally, a recurrent neural network is directly parameterized by the weight matrix $\mathbf{Q}$ and the bias term $\b$.
%Since in our problem we are interested in estimating components in $\mathbf{Q}$ and $\b$, we parameterize our network with waveform coefficients $\w$, and back propagate errors over only this component through the network. 
%This allows us to keep the known components fixed during the learning process and preserves the specific structure of our problem.  
%However, in our problem we are only interested in learning the unknown components of the network parameters. 
%In this case that is the waveform component of the forward model.
%Conventionally, parameterization of the network operator is through the network weights $\mathbf{Q}$ and bias $\b$. 
%The structure of our problem warrants a more specific parameterization. 
%Since we are merely interested in refining the waveform component of the forward model, we define the network by operator such that:
%Thus, we define the network operator as
%\begin{equation}\label{eq:networkParm}
%\mathcal{L}(\theta)[\d] = {\d}^* =  \text{diag}(\w) \tilde{\mathbf{F}} \hat{\brho}
%\end{equation}
%where $\mathcal{L}(\theta) : \Omega_{\d} \rightarrow {\Omega}^*_{\d}$ and is parameterized by the unknown $\theta = \{\w\}$.
%Furthermore, the final image representation or estimate ${\brho}^*$ is normalized as ${\brho}^* = \frac{\brho^L}{\| \brho^L\|_{\infty}}$ the final layer of the encoder (RNN) stage prior to backpropagation.
% $\brho^L$ denoting the representation produced by the final layer of the encoder (RNN) stage. 

In addition to the unknown waveform in the passive SAR forward model, the optimization hyperparameters can be also learned.
For our model, we include the threshold parameter of the network non-linearity, initialized as $\tau = \alpha \lambda$ into learning, and define network % the network is parameterized by 
parameter as $\theta = \{\w, \tau \}$.  
%\textcolor[rgb]{1.00,0.00,0.00}{\textbf{ Bariscan, should $\theta$ include the $\mathbf{K}$ and $\mathbf{Q}$. At this point it seems to change so I am unsure. Or should this $\theta$ be $\theta = \{ \theta, \w, \lambda \}$? If this is changed, be aware it needs to be modified throughout. }}

%\textbf{Bariscan, I commented out the paragraph that was here, it needs to be reworded, and is a discussion more appropriate for the conclusion}
%As DL framework connects to optimization framework at forward propagation by the design of the network, at backpropagation step network parameters are updated in a task specific manner. 
%Therefore, by parameterizing the network by the unknowns of forward propagation elements, backpropagation serves as a tool for refining these parameters such that the data estimates at forward propagation are improved. 
Given training data $D = \{\d_1, \d_2 \cdots \d_T\}$, we search a minimizer over the parameter space $\{\w, \tau \}$ for the following cost function:
\begin{equation}\label{eq:lossfunc_BP}
\mathcal{J}_D(\w, \tau) = \frac{1}{T} \sum_{n=1}^T \ell \left(  \text{diag}(\w) \tilde{\mathbf{F}} {\brho}^*_n, \d_n\right), 
\end{equation}
where $\ell$ is a properly chosen loss function. 
For  $\ell$, we pick the $\ell_2$-norm of the mismatch between the data synthesized by the network and the input data. 
Furthermore, we incorporate constraints into the cost function to enforce prior information on $\w, \tau$.
Most significantly, we focus our attention on constraining the underlying waveform by an \emph{a priori} functional form. 
For our problem, we consider that the unknown waveform coefficients are sampled from a flat spectrum signal, and restrict $\w$ to have unit modulo entries. 
For $\tau$, we invoke a non-negativity constraint, and formulate the backpropagation problem as
\begin{equation}
BP := \underset{\w, \tau}{\text{minimize}} \frac{1}{T} \sum_{n=1}^T  \| \text{diag}(\w) \tilde{\mathbf{F}} {\brho}^*_n - \d_n \|_2^2 + i_{C_{\w}} ( \w ) + i_{C_{\tau}}(\tau), 
\end{equation}
where $i_C( \cdot)$ denotes the indicator function of subscript set $C$, $C_{\w}$ and $C_{\tau}$ denote constraint sets on parameters $\w$ and $\tau$, respectively. 
The minimization is then performed by projected SGD.
Taking gradient steps over the smooth $\mathcal{J}_D$ term, and projecting onto constraint sets $C_{\w}$ and $C_{\tau}$, we obtain the updates as
\begin{equation}\label{eq:updateEq1}
\w^{l+1} = \mathcal{P}_{C_{\w}} \left(  \w^{l}  - \frac{\eta_{\w}}{|B|} \sum_{n \in I_B} \nabla_{\w} \ell (  \text{diag}(\w) \tilde{\mathbf{F}} {\brho}^*_n , \d_n ) \vert_{\theta = \theta^l} \right),
\end{equation}
\begin{equation}\label{eq:updateEq2}
\tau^{l+1} = \mathcal{P}_{C_{\tau}} \left( \tau^{l} - \frac{\eta_{\tau}}{|B|} \sum_{n \in I_B} \frac{\partial \ell (  \text{diag}(\w) \tilde{\mathbf{F}} {\brho}^*_n , \d_n )}{\partial \tau} \vert_{\theta = \theta^l} \right)
\end{equation}
where $\theta^l = \{\w^l, \tau^l\}$ denotes the parameter values at the $l^{th}$ iteration in backpropagation, $\eta_{\w}$ and $\eta_{\tau}$ are the learning rates (or step-size), $B \subseteq D$ is the randomly selected subset of the training data, $|B|$ is the cardinality of the subset $B$, $I_B$ is an index set corresponding to $B$.
$\mathcal{P}_C$ denotes the projection operator corresponding to the constraint set $C$ such that for the specified constraints for $\w$ and $\tau$, we have
\begin{equation}
\left( \mathcal{P}_{C_{\w}} ( \tilde{\w} ) \right)_i = \frac{\tilde{\w}_i}{| \tilde{\w_i} |} , \quad  \mathcal{P}_{C_{\tau}} ( \tilde{\tau} ) = \text{max} ( \tilde{\tau}, 0 ).
\end{equation}

As mentioned earlier, due to the nested nonlinear structure of the recurrent auto-encoder estimator, \eqref{eq:lossfunc_BP} is a highly non-convex optimization problem.
In addition, first order methods are prone to converging to local minima, which adds further difficulty to estimating  $\w$. 
However, our parameterization enforces the problem structure of SAR imaging, and places the network in a neighborhood over the loss surface only upto a diagonal multiplier of the true forward model, along with any prior knowledge of the functional form of $\w$.
Therefore, by the updates in \eqref{eq:updates}, a stationary point in the neighborhood of a strong initial point is searched.
%Hence, optim over the network parameters is likely to get stuck in sub-optimal stationary points of \eqref{eq:lossfunc_BP}. 
%However, since the model is initialized with the known components of the SAR forward model and a priori known or guessed functional form for $\w$, although likely sub-optimal, the stationary point that we obtain for waveform coefficients will be at a . 
%For good initialization, this already places us at a relatively low loss function value over the parameter space.
%This motivates the idea of using back propagation effectively as a tool for \emph{refining} the solver of the optimization problem of the forward propagation.
Hence, backpropagation is used as a tool for \emph{refining} the SAR forward model and improving image reconstruction implemented by forward propagation. 

%The minimization is then performed by stochastic gradient descent, a method in which an estimate of the loss function gradient is estimated for a randomly selected subset of the training data $B \subseteq D$.

%\textbf{Bariscan, the waveform will be complex and this will require taking Wirtinger derivatives. This should be mentioned.}
\subsection{Network Derivatives}
%The minimization is then performed by a projected SGD of the form:
%\begin{equation}\label{eq:updateEq1}
%\w^{l+1} = \mathcal{P}_{C_{\w}} \left(  \w^{l}  - \eta \frac{1}{|B|} \sum_{n \in I_B} \nabla_{\w} \ell \left(  \text{diag}(\w) \tilde{\mathbf{F}} {\brho}^*_n , \d_n\right) \vert_{\theta = \theta^l} \right)
%\end{equation}
%\begin{equation}\label{eq:updateEq2}
%\lambda^{l+1} = \mathcal{P}_{C_{\lambda}} \left( \lambda^{l} - \eta \frac{1}{|B|} \sum_{n \in I_B} \frac{d\ell \left(  \text{diag}(\w) \tilde{\mathbf{F}} {\brho}^*_n , \d_n\right)}{d\lambda} \vert_{\theta = \theta^l} \right),
%\end{equation}
%where $\theta^l = \{\w^l, \tau^l\}$ denotes the parameter values at the $l^{th}$ iterate in backpropagation, $\eta$ is the learning rate (or step-size), $B \subseteq D$ is the randomly selected subset of the training data, $|B|$ is the cardinality of the subset $B$, $I_B$ is an index set corresponding to $B$.
%$\mathcal{P}_C$ denotes the projection operator corresponding to the constraint set $C$, such that for the specified constraints for $\w$ and $\tau$ we have that:

An important consideration is the computation of the backpropagation equation for waveform coefficients. 
Since the objective being minimized is a real-valued function of a complex variable $\w$, we have 
%to use the definition of the complex gradient operator and 
to perform complex backpropagation on the parameters defined as %as derived for our recurrent-autoencoder model in  \cite{yonel2018deep} by
\begin{equation}\label{eq:ComGrad}
\nabla_{\w} \mathcal{\ell} = \overline{\left(\frac{\partial \ell}{\partial \w}\right)}, 
\end{equation}
where $\bar{(\cdot)}$ denotes complex conjugation and the partial derivative in \eqref{eq:ComGrad} is the  {Wirtinger derivative} \cite{kreutz2009complex}. %defined as
%is with respect to $w = w_R + j w_I$, $w_R, w_I \in \mathbb{R}$, denoting a scalar entry of $\w$, and
%\begin{equation}\label{eq:WirtDerivatives}
%\frac{\partial}{\partial w} = \frac{1}{2}\big( \frac{\partial}{\partial w_{R}}  - j \frac{\partial}{\partial w_{I}} \big) , \ \frac{\partial}{\partial {\bar{w}}} = \frac{1}{2}\big( \frac{\partial}{\partial w_{R}}  + j \frac{\partial}{\partial w_{I}} \big)
%\end{equation}
%where $w = w_R + j w_I$, $w_R, w_I \in \mathbb{R}$, denotes a scalar entry of $\mathbf{w}$.  
%\footnote{(\ref{eq:WirtDerivatives}) reduces to complex differentiation for holomorphic functions.}
%The Wirtinger derivative relies on a new interpretation of the complex plane, where a variable and its conjugate are decoupled such that the partial derivative of $w \in \mathbb{C}$ with respect to its conjugate is $0$.
%This interpretation allows an approach for real valued, or in general non-holomorphic functions of complex variables in optimization.
%For real valued functions, the gradient becomes identical to the partial derivative of the function with respect to the conjugate term, yielding conjugate updates for $w$ and $\bar{w}$ variables \cite{candes2015phase}.
%Hence optimization can be performed only over the variable $w$ for which the function is holomorphic given $\bar{w}$. 

Notably, $\w$ parameterizes both $\mathbf{Q}$ and $\mathbf{F} = \text{diag}(\w) \tilde{\mathbf{F}}$. 
Writing the partial derivative of the loss function $\ell$ with respect to $\w$, from the chain rule we have
\begin{equation}\label{eq:wderiv}
\frac{\partial \ell(\d^*, \d)}{\partial \w} = \frac{\partial \mathbf{Q}}{\partial \w} \frac{\partial \ell(\d^*, \d)}{\partial \mathbf{Q}} + \frac{\partial \mathbf{F}}{\partial \w} \frac{\partial \ell(\d^*, \d)}{\partial \mathbf{F}}.
\end{equation}
Hence, we compute the derivative with respect to $\w$  by first computing derivatives with respect to $\{\mathbf{Q}, \mathbf{F}\}$ as
\begin{eqnarray}\label{eq:FinalNetBackprop}
\frac{\partial \ell(\d^*, \d)}{\partial \mathbf{Q}} & = & \frac{\partial \brho^*}{\partial \mathbf{Q}} \frac{\partial \ell(\d^*, \d) }{\partial {\brho}^*},    \nonumber \\
\frac{\partial \ell(\d^*, \d)}{\partial \mathbf{F}} & = & \frac{\partial {\brho}^*}{\partial \mathbf{F}} \frac{\partial \ell(\d^*, \d) }{\partial {\brho}^*}  + (\frac{\partial \mathbf{F}}{\partial \mathbf{F}} \brho^*) (\bar{\d^*} - \bar{\d}).
\end{eqnarray}
The derivative with respect to $\tau$ is simply the real valued partial derivative of the form %\footnote{Both partial derivatives in the chain rule of \eqref{eq:tauderiv} are real valued, due to having a real-valued loss function, image estimate and variable.} partial derivative of the form
\begin{equation}\label{eq:tauderiv}
\frac{\partial \ell(\d^*, \d)}{\partial \tau}  =  \frac{\partial {\brho}^*}{\partial \tau}  \frac{\partial \ell(\d^*, \d) }{\partial {\brho}^*}.
\end{equation}

The second component of the derivative with respect to  $\mathbf{F}$ in \eqref{eq:FinalNetBackprop} originates from having the linear decoder stage that projects the image estimate $\brho^*$ on the received signal space by $\mathbf{F}$. 
In \eqref{eq:FinalNetBackprop}, the partial derivative of $\mathbf{F}$ with respect to itself yields an identity {tensor} of size $M \times N \times M \times N$, which multiplies the image estimate $\brho^* \in \mathbb{R}^N$ to yield an $M \times N \times M$ tensor. %The corresponding derivations are detailed in Appendix \ref{sec:App1}. 
We provide a detailed derivation of network derivatives in Appendix \ref{sec:App1}.
%Due to having real valued representations such that $\brho^k \in \mathbb{R}^M$, and $\bar{\brho}^* = {\brho}^*$ the complex backpropagation equation becomes:
%\begin{equation}
%\frac{\partial \ell(\d^*, \d) }{\partial {\brho}^*}  = \big( \mathbf{F}^T (\bar{\d^*} - \bar{\d}) + \mathbf{F}^H ({\d}^* - \d) \big)
%\end{equation}s

To backpropagate through the recurrent encoder component, we use the backpropagation through time algorithm (BPTT). 
Since the parameters $\{\mathbf{Q}, \mathbf{F}, \tau \}$ are shared among layers, and only the error resulting from the final layer is considered in the optimization, %over the parameters in \eqref{eq:lossfunc_BP}, 
BPTT computes the derivatives in the RNN-encoder as
\begin{equation}\label{eq:BPTTRnn}
\frac{\partial \brho^*}{\partial \theta} = \left(\sum_{i = 1}^{L} \frac{\partial \brho^k}{\partial \theta} \frac{\partial \brho^L}{\partial \brho^k}\right) \frac{\partial {\brho}^*}{\partial \brho^L},
\end{equation}
%where $\d^* = \text{diag}(\w) \tilde{\mathbf{F}} {\brho}^*$ is the synthesized data, and 
%where $\brho^* = \brho^*(\theta, \d)$ is the final image estimate of the network for input data sample $\d$, 
%and $\theta$ is a surrogate for network parameters $\{\mathbf{Q}, \mathbf{F}, \tau \}$, 
where  $\brho^k$, $k = 1, \cdots L$ is the network representation generated at the $k^{th}$ layer. 

%Since the $\brho^*$ is simply the normalized image obtained from the final layer of the RNN-encoder, we deploy another chain rule for the normalization stage:
%\begin{equation}\label{eq:DefNablRhostar}
%\frac{\partial {\brho}^*}{\partial \brho^L}\frac{\partial \ell(\d^*, \d) }{\partial {\brho}^*}= \big(- \frac{1}{\|{\brho}^L \|_{\infty}^2} \frac{\partial \|{\brho}^L\|_{\infty}}{\partial {\brho}^L} {\brho^L}^T $$ 
%$$+ \frac{1}{\|{\brho}^L\|_{\infty}} \mathbf{I}_{M\times M}\big) \times 2 \text{ Re}\{\mathbf{F}^H ({\mathbf{d}}^* - \mathbf{d})\}
%\end{equation}
%The first term of the multiplication is derived from the normalization derivative $\frac{\partial {\brho}^*}{\partial {\brho}^L}$. 

%To conclude with our final update terms, further details on derivations are provided in the Appendix \ref{sec:}. 
Despite the general form we present in this section, a distinct effect of the flat spectrum constraint we place on $\w$ for signals such as QPSK or OFDM can be observed at backpropagation. 
%Specifically, by setting initial $\w^0$ as a unit modulo entry signal, and projecting $\w$ iterates onto the unit sphere in $\mathbb{C}^M$ by \eqref{eq:updateEq1}, the image domain filter $\mathbf{Q}$ becomes independent of the parameter $\w$. 
As explicitly shown in \eqref{eq:networkArch}, $\mathbf{Q}$ only has dependence on $\w$ through its elementwise modulus by the $\text{diag}(| \w |^2)$ term. 
By setting the initial $\w^0$ as a unit modulo entry signal, and projecting $\w$ iterates onto the unit sphere in $\mathbb{C}^M$ by \eqref{eq:updateEq1}, $\text{diag}(| \w |^2)$ is merely the identity matrix and $\mathbf{Q}$ is effectively fixed through the training procedure. 
Therefore its dependence on the network parameterization is dropped. 
%and its contribution to the partial derivative with respect to $\w$ in \eqref{eq:wderiv} is dropped. 
Removing the contribution of $\mathbf{Q}$ to the partial derivative in \eqref{eq:wderiv} yields the following final update form for the parameter $\w$: 
\begin{equation}\label{eq:wderiv_fin}
\frac{\partial \ell(\d^*, \d)}{\partial \w} = \frac{\partial \mathbf{F}}{\partial \w} \left[ \left(\sum_{i = 1}^{L} \frac{\partial \brho^k}{\partial \mathbf{F}} \frac{\partial \brho^L}{\partial \brho^k}\right) \frac{\partial {\brho}^*}{\partial \brho^L}\frac{\partial \ell(\d^*, \d) }{\partial {\brho}^*} + (\frac{\partial \mathbf{F}}{\partial \mathbf{F}} \brho^* )(\bar{\d^*} - \bar{\d}) \right].
\end{equation}
 
%As mentioned earlier, due to the nested non-linear structure of the estimate generated by the RNN-encoder, \eqref{eq:lossfunc_BP} is a highly non-convex, high dimensional, optimization problem, with the added difficulty that SGD is prone to converging to local minima.
%However, our parameterization enforces the problem structure of SAR imaging, and places the network in a neighborhood over the loss surface only upto a diagonal multiplier on the true forward model, along with any prior knowledge of the functional form of $\w$.
%In \eqref{eq:updates}, SGD searches for a stationary point in the neighborhood of a strong initial point.
%%Hence, optim over the network parameters is likely to get stuck in sub-optimal stationary points of \eqref{eq:lossfunc_BP}. 
%%However, since the model is initialized with the known components of the SAR forward model and a priori known or guessed functional form for $\w$, although likely sub-optimal, the stationary point that we obtain for waveform coefficients will be at a . 
%%For good initialization, this already places us at a relatively low loss function value over the parameter space.
%%This motivates the idea of using back propagation effectively as a tool for \emph{refining} the solver of the optimization problem of the forward propagation.
%Hence, back propagation is used as a tool for \emph{refining} the SAR forward model and improving the solver implemented during forward propagation. 
%\subsection{Computational Complexity}

\begin{figure}
  \centering
  \includegraphics[width=0.37\textwidth]{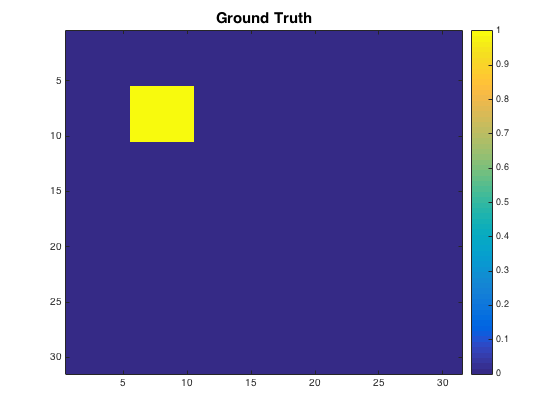}
   \includegraphics[width=0.37\textwidth]{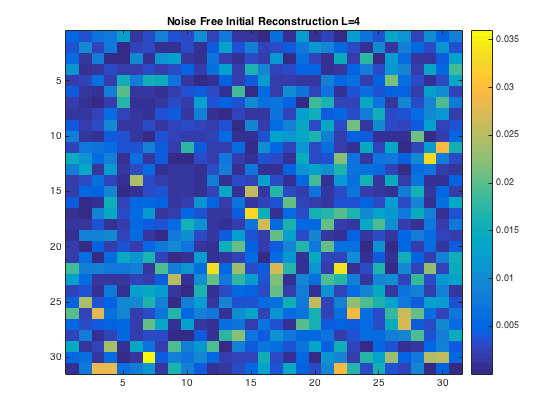}
    \caption{\emph{Ground truth image used in experiments (left) vs. the image reconstructed by the RNN encoder with the randomly initialized waveform coefficients (right).} The target is completely lost and suppressed in background noise due to the mismatch introduced by the wrong waveform in reconstruction. }
 \label{fig:Figure1}
\end{figure}

\section{Numerical Simulations}\label{sec:simulation}

%\subsection{Experimental Setup}
%We present numerical simulations to demonstrate the performance of DL-based simultaneous waveform estimation and image reconstruction. 
\subsection{Scene and Imaging Parameters}

We assume isotropic transmit and receive antennas, and simulate a transmitted signal with bandwidth and center frequency of $8$MHz and $760$MHz, respectively.
The simulated waveform is modulated using QPSK, in which the symbols are generated from an i.i.d. uniform distribution. 
This corresponds to approximately $20$m range resolution for monostatic SAR.
Thus, we simulate a $620\times 620$ m$^2$ scene and discretize it into $31\times 31$ pixels as shown in Figure \ref{fig:Figure1}.
% with the origin of the coordinate system located at the center of the scene at pixel $(16,16)$. 

The receiver antenna traverses a circular trajectory, defined as $\bgamma_R(s) = [7 \cos(s), 7 \sin(s), 6.5]$ km.
The transmitter is fixed and located at $\bgamma_T = [11.2, 11.2, 0.2]$ km. 
 %described in Section \ref{Sec:SARDL}.
The aperture is sampled uniformly into $128$ uniform samples, and the bandwidth is sampled uniformly into $64$ samples.

\subsection{Training and Testing Sets}\label{sec:Train}

We generate training samples consisting of randomly generated sparse scenes with a single point or extended target that varies in rectangular shape and location. 
The length and width of each rectangular target are sampled from a uniform distribution in the range [1, 6]. %$\times$ [1, 6] pixels. 
The targets are placed randomly within the range of [3, 28] $\times$ [3, 28] pixels. 
We then generate received SAR signals for each scene using the full forward model in \eqref{eq:datamodel3} described in Section \ref{sec3}, and add a realization of white Gaussian noise vector on each SAR measurement we've generated. 
%We add a realization of white Gaussian noise on the received SAR signals. 
We consider $6$ levels of SNR, $-20$, $-15$, $-10$, $-5$, $0$ and $10$ dB, and create a training set of noisy received SAR signals of randomly generated scenes at each SNR level, to form $6$ statistically independent sets. 
The proposed model is trained using the $6$ training sets separately to evaluate the robustness of the model to different noise levels.   

In testing, we use measurements collected from a single scene of interest. The test scene is displayed in Figure \ref{fig:Figure1} and the backscattered field is generated by the forward model in \eqref{eq:datamodel3}. 
20 different realizations of white Gaussian noise at $-20$, $-15$, $-10$, $-5$, $0$ and $10$ dB SNR are used to form $6$ sets, each consisting of 20 samples of measurements. %, hence each consisting of 20 samples of SAR measurements. 
Each test set is fed into the model trained with the corresponding SNR level of received SAR signals. 
We evaluate the reconstruction performance as the average over 20 results for statistical accuracy.

%In each training set, we add a realization of white Gaussian noise at specified SNR level to each sample. 
%The model is trained using each training set separately to test the robustness under different noise levels.   
%We consider SNR levels of $-20$, $-15$, $-10$, $-5$, $0$ and $10$ dB in SAR measurements collected from randomly generated scenes.

For data collection, we envision a two-stage protocol to form training and test sets as proposed in \cite{yonel2018deep}. 
In the first stage, an airborne receiver collects test data from a scene of interest. 
In the second stage, arbitrary reflectors are placed in the scene to form either extended or point targets and training data is collected under the same imaging geometry as before.
%Another protocol is to collect the training data set over
%the course of an extended period during which temporary
%structures may appear creating perturbations in the background
%scene of interest. 
%Note that neither the location nor the shape of the foreground scatterers need to be known in order to use the training data in our unsupervised training scheme.
In the data collection procedure, we make the assumption that changes on the transmitted signal are negligible in slow-time. 
%This assumption allows the waveform coefficients of the transmitted signal to be shared among the training samples and the testing data, such that learning is consistent with our formulation. 
Although our formulation does not require a slow-time stationarity, collection of a training set under a slow time varying waveform is a complication that has to be alleviated, and is the main focus of our future work. 

%\subsection{Test Data}
%For our training set, we generate sparse scenes with a single extended target of varying rectangular shapes and location.
%Length and width of each rectangular target is chosen randomly and lies in the range $[1,6]\times[1,6]$ pixels. 
%The targets are placed randomly within the range of $[3,28]\times[3,28]$ pixels.
%We then generate SAR data for each image using the true waveform with the model in equation \eqref{eq:datamodel3}. 
%Test sets are formed from the scene of interest with $20$ realizations of additive white Gaussian noise on the backscattered field at different dB levels. 
%The model is trained and tested using data collected from the same noise level in each experiment, varied from $-10$, $0$, $10$ and $20$ dB noise on the SAR measurements. 

\subsection{Network Design and Initialization}

We implement the proposed network with an RNN-encoder of $L = 4$ layers, and the phaseless soft-thresholding activation function introduced in Section \ref{sec4}. 
The model is trained for $10$ epochs for each experiment. %although we experiment with $15$ for $-20$ and $-15$ dB SNR cases. 
%Despite describing the training in the general SGD setting, 
We limit the number of training samples based on the results of our previous study \cite{yonel2018deep} and set it to $10$.
%This is motivated to increase practicality of our approach under the 
We perform batch gradient descent, which corresponds to a single parameter update per epoch. 
%Due to opting for limited number of training samples under the indication of our previous study \cite{yonel2017deep}, we perform batch gradient descent, corresponding to a single parameter update per epoch.  
%We use an adaptive learning rate of $\eta_l = \eta_l /(1 + l)$ in training, where $l = 1,\dots$ is the index of the epoch, with an initial step size of $\eta_0 = 1e-4$. 
We set the learning rate as $\eta_{\w} = 10^{-4}$ for waveform coefficients, and as $\eta_{\tau} = 10^{-6}$ for the threshold parameter.  %for $-15$ and $-20$ dB cases, $1e-7$ for the others. 

The network weight matrix and bias terms are initialized with the known components of the forward model ${\mathbf{F}}$ in \eqref{eq:networkParm}. 
We set the initial regularization parameter as $\lambda = 10$, and set $\alpha = 1e-5$, upper bounded by the reciprocal of the largest eigenvalue of $\tilde{\mathbf{F}}^H \tilde{\mathbf{F}}$. 
%To keep the number of unknowns at a reasonable level, we assume that waveforms are static with respect to the slow-time variable. 
Accordingly to the discussion in Section \ref{sec:Train}, we assume that transmitted waveforms are static with respect to the slow-time variable. 
As demonstrated in Section \ref{sec5}, we assume \emph{a priori} knowledge is available, and constrain the parameter $\mathbf{w}$ using the QPSK properties, and project the entries of $\w$ at each update onto the unit circle in $\mathbb{C}$, as in \eqref{eq:updateEq1}. 
%For our demos, we $\w$ is further constrained by computing derivatives generated from a fixed waveform over slow-time variables to decrease the number of unknown waveform coefficients to only the number of fast-time samples. 
%A random waveform subject to the identical constraints is generated as initialization of the parameter $\w$. 
We initialize $\w$ by the real-valued flat spectrum signal of all ones instead of random initialization to standardize our evaluation of different experiments. 

\begin{figure}
  \centering
   \includegraphics[width=0.37\textwidth]{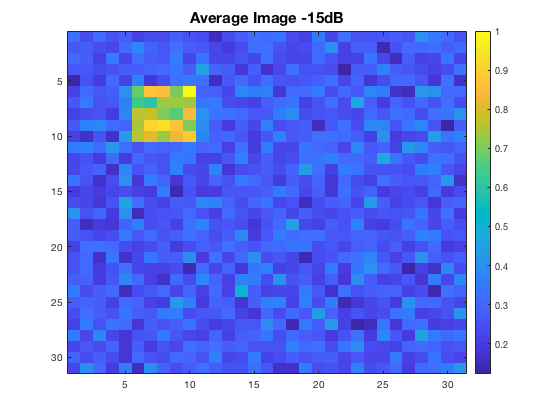}
  \includegraphics[width=0.37\textwidth]{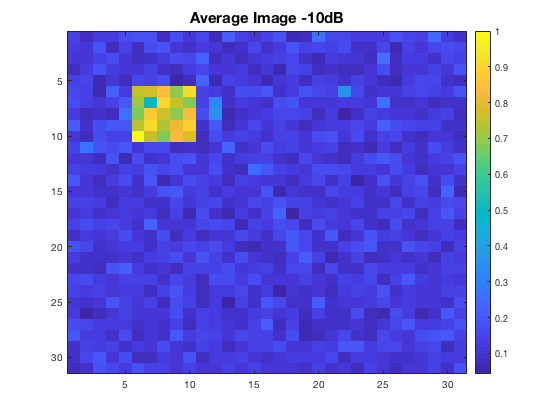}
  \includegraphics[width=0.37\textwidth]{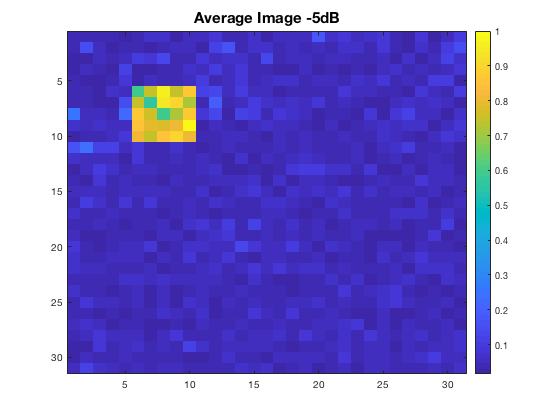}
  \includegraphics[width=0.37\textwidth]{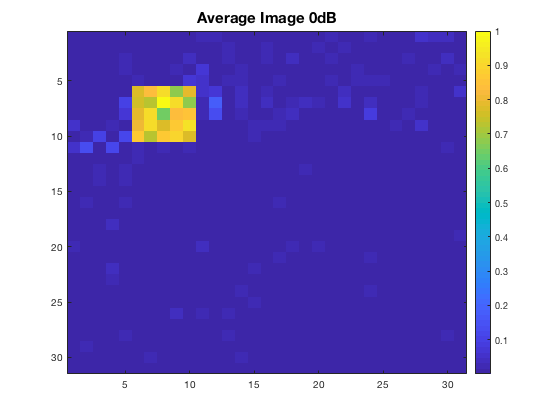}
    \caption{\emph{Reconstructed average images by the proposed model under -15 dB, -10 dB, -5 dB and 0 dB SNR levels at data collection for training and test sets.} Each image is formed by averaging the 20 test samples under different realizations at the same SNR. The model has learned suitable parameters such that imaging performance is drastically improved over the initialization image for every noise level under consideration. } 
 \label{fig:Figure3}
\end{figure}

\subsection{Performance Evaluation}\label{sec:PerfEval}

We evaluate the reconstruction performance on the image and waveform coefficients using the following figures of merit:
\begin{equation}\label{eq:dataMismatch}
L_{\d}( \mathcal{\brho^*}^l )= \frac{\| \text{diag} (\mathbf{w}^l) \tilde{\F} (\mathcal{\brho^*})^l - \d \|_2^2}{\| \d \|_2^2}, \ L_{\brho}( \mathcal{\brho^*}) = \frac{\| \mathcal{\brho^*} - \brho \|_2^2}{\| \brho \|_2^2},
\end{equation}
measures the normalized data domain mismatch and the image domain error of the reconstructed image with respect to the ground truth SAR measurement and testing image, respectively, whereas
\begin{equation}\label{eq:Merits_img}
 C_{\brho}({\brho^*}) = \frac{|\text{E}[{\brho^*}_f] - \text{E}[\brho^*_b]|^2}{\text{var}[{\brho^*}_b] }, \  L_{\mathbf{w}_t} ( \mathbf{w}^l ) = \frac{\| \mathbf{w}_t - \mathbf{w}^l \|^2  } {\| \mathbf{w}_t \|^2},
\end{equation}
measure the contrast in the reconstructed image and the normalized waveform mismatch with respect to the ground truth QPSK coefficients, respectively. 
%\begin{equation}\label{eq:Merit_wform}
% L_{\mathbf{w}_t} ( \mathbf{w}^l ) = \frac{\| \mathbf{w}_t - \mathbf{w}^l \|^2  } {\| \mathbf{w}_t \|^2},
%\end{equation}
%measures the normalized waveform mismatch with respect to the ground truth QPSK coefficients. 
$(\brho^*)^l$ denotes the normalized image generated by the RNN encoder, with the parameters obtained at epoch $l$, and $\brho$ is the ground truth image.
$\mathbf{w}_t$ is the ground truth waveform coefficient vector and $ \mathbf{w}^l$ is the learned waveform coefficient vector at epoch $l$. $\brho^*_{f}$ and $\brho^*_{b}$ are the foreground and background components of the reconstructed image, respectively. 
$\text{E}[\cdot]$ stands for statistical expectation, $\text{var}[\cdot]$ stands for statistical variance and $\d$ is the input SAR data.
%The figures of merit are averaged over the $20$ images reconstructed from different realizations of noise at the same dB level.
Curves corresponding to each SNR value demonstrate the performance of the model trained on measurements corresponding to that noise level, and evaluated on test samples collected at the same SNR. 

To evaluate the resolution improvement achieved by learning, we examine the bias terms of the DL-based model.
This corresponds to evaluation of backprojection reconstruction following matched filtering with the learned waveform coefficients. 
We compare the peak and average background of the reconstructed image with learned waveform coefficients to the one obtained with the true waveform, and to that of reconstruction by backprojection without matched filtering.
Essentially, the last case is equivalent to the initial waveform set to all $1$'s in its bandwidth.
Evaluation is performed on the phantom displayed in Figure \ref{fig:8a}, with respect to how the two point targets are resolved in horizontal and vertical directions, as well as how the weak point target in (12,17) is resolved from the background noise in our evaluations.

%\begin{figure}
%\centering
%  \includegraphics[width=0.38\textwidth]{org_wform}
%   \caption{\emph{Linearly reconstructed phantom image by first match filtering with the true underlying waveform coefficients, followed by backprojection, at $-10$ dB SNR.}}
% \label{fig:Figure6}
%\end{figure}

%\begin{figure}
%  \centering
%  \includegraphics[width=0.38\textwidth]{-10BP}
%  \caption{\emph{Linearly reconstructed phantom image from directly backprojecting data at $-10$ dB SNR.}} %The targets are nearly indistinguishable without matching the measurements with a highly correlated estimate to the underlying waveform.}
%  \label{fig:Figure8a}
%\end{figure}

%\begin{figure}
%\centering
%  \includegraphics[width=0.38\textwidth]{learn_wform}
%  \caption{\emph{Linearly reconstructed phantom image by first match filtering with the learned waveform coefficients, followed by backprojection, at $-10$ dB SNR.}}
%   \label{fig:Figure8}
%\end{figure}
\begin{figure}
  \centering
  \includegraphics[width=0.37\textwidth]{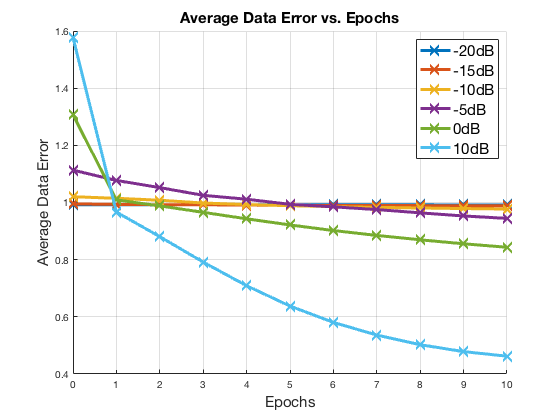}
  \includegraphics[width=0.37\textwidth]{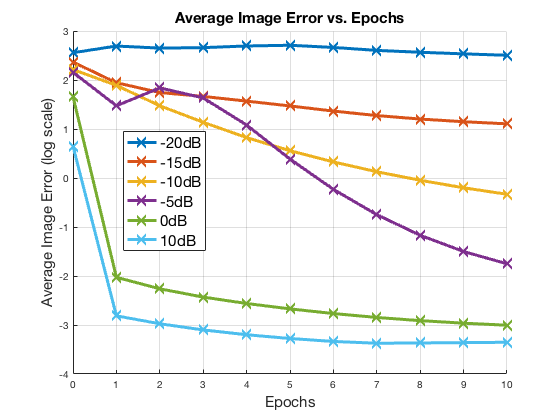}
  \includegraphics[width=0.37\textwidth]{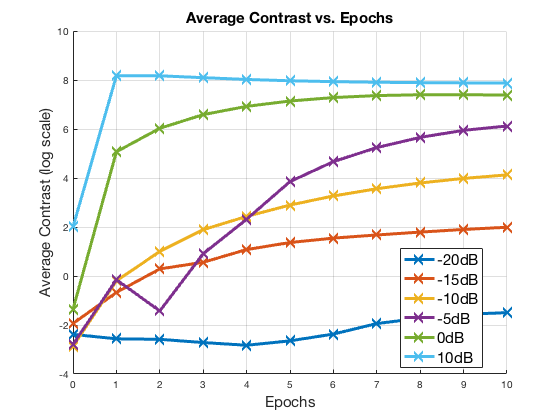}
  \includegraphics[width=0.37\textwidth]{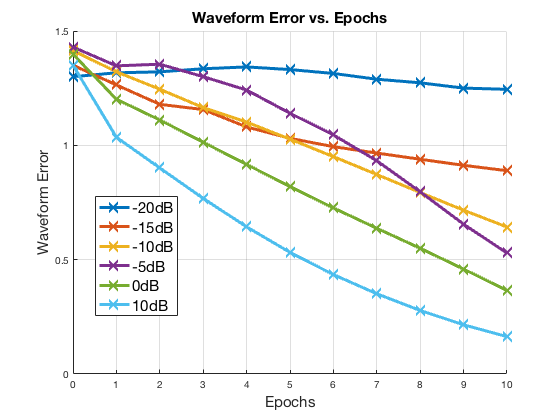}
\caption{\emph{Average values of the proposed data mismatch, image mismatch \eqref{eq:dataMismatch}, image contrast and waveform error metrics \eqref{eq:Merits_img} over $20$ test samples at $-20$, $-15$, $-10$, $-5$, $0$ and $10$ dB SNR cases.} Curves are obtained from reconstruction results with the parameter estimates generated by the network at each epoch, trained with SAR measurements under corresponding SNR.}
 \label{fig:Figure4}
\end{figure}

\subsection{Results}

%
%\begin{figure*}
%  \centering
%  \includegraphics[width=0.32\textwidth]{hor15}
%  \includegraphics[width=0.32\textwidth]{hor17}
%  \includegraphics[width=0.32\textwidth]{ver12}
%  \caption{\emph{Cross sections of linearly reconstructed images at $-10$ dB SNR in Figures \ref{fig:Figure2}, and \ref{fig:Figure5}.} The background pixels at the cross section are averaged to depict the noise level with respect to peak values. All values displayed in logarithmic scale.}
%  \label{fig:Figure6}
%\end{figure*}

Our simulations show that the DL-based approach achieves accurate reconstruction performance under SNR scenarios above $-20$ dB for all metrics under consideration.
To display the performance visually, we present the reconstructed images by the model under SNR levels of $-15$, $-10$, $-5$, and $0$ dB in Figure \ref{fig:Figure3}. 
The images displayed in the figure demonstrate the impact of waveform and threshold learning by the DL-based method as described in Section \ref{sec:ForwProp}. 
By fixing the image-domain filters due to the constraints on $\w$, learning the waveform coefficients become equivalent to refining the backprojection image. 
Hence, waveform learning directly impacts the placement and strength of target pixels, whereas threshold learning controls the amount of background suppression in the image. 
%The impact of threshold learning is most visible in the $-15$ dB image, as the threshold learning rate higher than the other experiments yields similar background suppression while trading off gradual suppression of targets pixels. 
It is observed in Figure \ref{fig:Figure3} that with the exception of the $-20$ dB case, clear background suppression and geometric fidelity of the extended target are achieved by our method despite initializing the model with a waveform that has poor correlation to the true one. 
Moreover, image contrast and image mismatch metrics, as well as the decay in waveform error shown in Figure \ref{fig:Figure4} validate the main arguments of our approach, as the waveform coefficients are learned to the extent of high correlation with the underlying QPSK signal such that the model produces enhanced imagery. 

As expected, the performance of the method degrades gracefully as the noise level increases both in image reconstruction and waveform estimation. 
For the $-20$ SNR case, the gradual improvement in the waveform is insufficient to impact reconstruction performance, as indicated by negligible changes in image domain metrics as shown in Figure \ref{fig:Figure4}. 
However, the drastic impact of waveform estimation can be observed in the $-10$ dB case. 
Despite no indicative improvement on the data mismatch function over epochs similar to the $-20$ dB case, the algorithm learns a much more correlated waveform coefficients, which improves the reconstruction performance significantly. 

\begin{figure}
\centering
\begin{subfigure}[t]{\textwidth}
\centering
\includegraphics[width=0.32\textwidth]{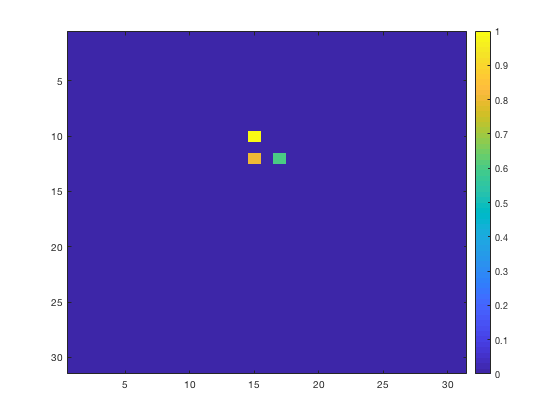}
\includegraphics[width=0.32\textwidth]{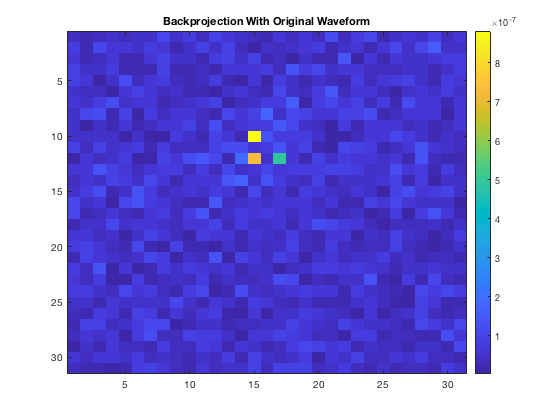}
\includegraphics[width=0.32\textwidth]{org_wform}
\caption{}
\label{fig:8a}
\end{subfigure}
\begin{subfigure}[b]{\textwidth}
\centering
\includegraphics[width=0.32\textwidth]{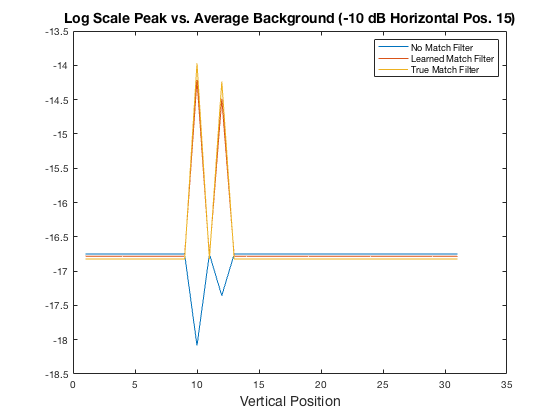}
\includegraphics[width=0.32\textwidth]{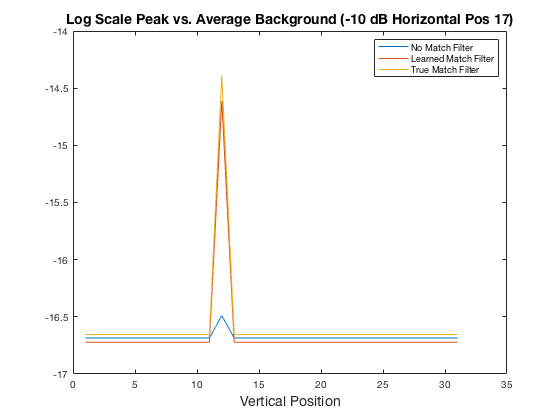}
\includegraphics[width=0.32\textwidth]{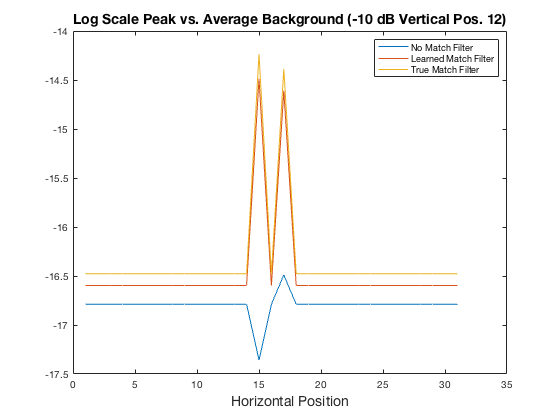}
\caption{}
\label{fig:8b}
\end{subfigure}
\caption{(\ref{fig:8a}) \emph{Phantom used to evaluate resolution, and linearly reconstructed images by the true, and learned waveform coefficients.} The point targets are located in range bins $15$ and $17$, cross range bins $10$ and $12$. (\ref{fig:8b}) \emph{Cross sections of linearly reconstructed images at $-10$ dB SNR in \ref{fig:8a}.} The background pixels at the cross section are averaged to depict the noise level with respect to peak values. All values displayed in logarithmic scale.}
\label{fig:Figure8}
\end{figure}

We demonstrate the resolution performance of the DL-based method for the $-10$ dB case as discussed in Section \ref{sec:PerfEval}. 
%The images reconstructed by backprojection after matched filtering with true, initial and learned waveform coefficients are provided in Figures \ref{fig:Figure6}, \ref{fig:Figure8a}, and \ref{fig:Figure8}, respectively. 
The images reconstructed by backprojection after matched filtering with true, and learned waveform coefficients are provided in Figure \ref{fig:8a}.  
%Figures \ref{fig:Figure6}, and \ref{fig:Figure8}, respectively. 
%and learned waveform coefficients are provided in Figures \ref{fig:Figure6}, and \ref{fig:Figure8}, respectively.
%Comparing Figures \ref{fig:Figure8a} and \ref{fig:Figure8}, the reconstruction using the waveform coefficients learned by the DL-based approach is superior in recovering and resolving point targets. 
From Figures \ref{fig:8a}, we see that the linear reconstruction using the learned waveform produces a nearly identical image as the one produced using the true waveform. 
This can also be observed in the cross-sections at horizontal and vertical directions that contain target pixels, which are provided in Figure \ref{fig:8b}, in $\log$ scale. %for reconstruction by match-filtering with true, initial, and learned waveform coefficients, respectively. 
The peak and average background values at each cross-section of the reconstructed image using learned waveform coefficients are highly consistent with the ones obtained with true waveform coefficients. %despite normalized mismatch error of $0.5$ at the end of epoch $10$ as indicated by Figure \ref{fig:Figure4}. 
Notably, the accuracy in image reconstruction is obtained despite a final normalized waveform mismatch of $0.5$ as shown in Figure \ref{fig:Figure4}, which indicates robustness of the method to errors in estimation. 

Overall, it can be presumed from our experiments that learning a sufficiently correlated waveform produces improved imagery.
The model offers considerable robustness to measurement noise even with limited number of training samples, which increases the applicability of our method in real-world scenarios. 
However, the limited performance in the $-20$ dB scenario can be traced to the limited number of samples used in training.
The poor improvement of waveform error suggests the gradient estimates are highly contaminated by noise, which can be avoided by averaging over more samples.
Handling such high noise scenarios may require accurate initialization, or more structural constraints on the functional form of the waveform, as well as increasing the number of training samples.
 
\section{Conclusion}\label{sec:conclusion}

This paper presented a novel deep learning based approach for simultaneous estimation of the scene reflectivity and the transmitted waveform. 
Our method requires a single receiver, providing reduced cost and improved simplicity over existing methods such as PCL and TDOA/FDOA backprojection.
We consider a passive imaging scenario in which the transmitter location is known, but the transmitted waveform is unknown.
We approach image reconstruction in a Bayesian framework and set up an optimization problem to estimate the scene reflectivity.
We formulate a proximal gradient descent algorithm to solve for the scene reflectivity, which we unfold for a fixed number of iterations to formulate an RNN parameterized by waveform coefficients. 
Hence, for a given waveform coefficient vector, the RNN becomes a solver for the scene reflectivity at forward propagation, and waveform coefficients become parameters that can be estimated by backpropagation. 

We then cascade the RNN with a decoding layer consisting of normalization and a linear forward map that synthesizes SAR measurements from the reconstructed scene reflectivities resulting in a recurrent auto-encoder architecture.
Thereby, we learn the transmitted waveform in an unsupervised manner by minimizing the mismatch between a set of received SAR signals and corresponding SAR measurements synthesized by the network. 
At backpropagation, we employ a flat spectrum constraint on the waveform by performing updates via projected stochastic gradient descent. 
Our formulation has applicability to wide range of spread spectrum signals that are common to transmitters of opportunity. 
The main advantage of our method is that the waveform estimation is performed in a task driven manner.
The DL-based model ultimately estimates waveform coefficients with the goal of producing accurate imagery by forward propagation. 
Moreover, the structural form for the transmitted signals is merely used as a prior by the means of a constraint set in our framework, and lack thereof does not limit our framework. 

We demonstrate the performance of our deep learning approach with numerical simulations, showing that with a limited number of training samples collected at realistic SNR levels, the model estimates QPSK modulated signals in a manner that produces accurate SAR imagery.
Furthermore, we show that our DL-based method is robust to estimation errors, as it reconstructs images highly consistent with the ones obtained by the true underlying waveform even in the presence of a non-negligible mismatch in learned and true waveform coefficients. 
In the future, we will pursue bypassing the slow-time stationarity assumption of transmitted waveform in training data collection, and explore decoding changing waveforms by our DL framework. 
Furthermore we will further pursue improving the performance of our method in low SNR scenarios, and test the estimation quality with other waveforms such as OFDM signals. 
 
\section*{Acknowledgement}
{
%This work was supported by the Air Force Office of
%Scientific Research (AFOSR) under the agreements under
%the agreement FA9550-16-1-0234, and the Naval Research
%Laboratory.
This work was supported by the Air Force Office of Scientific Research (AFOSR) under the agreement FA9550-16-1-0234, Office of Naval Research (ONR) under the agreement N0001418-1-2068 and by the National Science Foundation (NSF) under Grant No ECCS-1809234. 
E. Mason is supported by the Naval Research Laboratory.
}
%\begin{figure}[H]
%  \centering
%  \includegraphics[width=0.4\textwidth]{average_Contrasts}
%  \caption{\emph{Image contrasts.} Contrast vs. number of epochs, averaged over 20 test samples plotted in $\log$ scale. }\label{fig:true}
%\end{figure}

\bibliographystyle{IEEEtran}
\bibliography{ref_eric_new}

\appendix 
\section{Appendices}
\label{Sec:Appendix}
\subsection{Waveform Derivative} \label{sec:App1}
Due to having real-valued representations such that $\brho^k \in \mathbb{R}^N$, and $\bar{\brho}^* = {\brho}^*$ the first component of complex backpropagation equation becomes:
\begin{equation} \label{eq:lossfuncderiv}
\frac{\partial \ell(\d^*, \d) }{\partial {\brho}^*}  =  \mathbf{F}^T (\bar{\d^*} - \bar{\d}) + \mathbf{F}^H ({\d}^* - \d)  \\
= 2 \ \text{Re} \big(  \mathbf{F}^H ({\d}^* - \d) \big), 
\end{equation}
which is purely real valued as expected. 
%Since the $\brho^*$ is the normalized image obtained from the final layer of the RNN-encoder, we obtain the non-layer dependent component of the network derivative in \eqref{eq:wderiv_fin} 
From the chain rule with \eqref{eq:lossfuncderiv} and the normalization derivative $\frac{\partial {\brho}^*}{\partial {\brho}^L}$, we obtain the non-layer dependent component of the network derivative in \eqref{eq:wderiv_fin} by multiplying \eqref{eq:lossfuncderiv} with:
\begin{equation}\label{eq:DefNablRhostar}
\frac{\partial {\brho}^*}{\partial \brho^L} = \big(- \frac{1}{\|{\brho}^L \|_{\infty}^2} \frac{\partial \|{\brho}^L\|_{\infty}}{\partial {\brho}^L} {\brho^L}^T + \frac{1}{\|{\brho}^L\|_{\infty}} \mathbf{I}_{N\times N}\big),
\end{equation}
%\begin{equation}\label{eq:DefNablRhostar}
%\frac{\partial {\brho}^*}{\partial \brho^L}\frac{\partial \ell(\d^*, \d) }{\partial {\brho}^*}= \big(- \frac{1}{\|{\brho}^L \|_{\infty}^2} \frac{\partial \|{\brho}^L\|_{\infty}}{\partial {\brho}^L} {\brho^L}^T + \frac{1}{\|{\brho}^L\|_{\infty}} \mathbf{I}_{N\times N}\big) $$
%$$
%\times 2 \text{ Re}\{\mathbf{F}^H ({\mathbf{d}}^* - \mathbf{d})\},
%\end{equation}
%The first term of the multiplication is derived from the normalization derivative $\frac{\partial {\brho}^*}{\partial {\brho}^L}$. 
where the partial of the infinity norm of $\brho^L$ is simply a column vector with entry $1$ at the index of the maximal element of $\brho^L$, and $0$'s elsewhere. 
%Computing the second expression from the product rule in equation \eqref{eq:wderiv_fin}, first we evaluate a $4D$ tensor - vector multiplication. 

First we consider the second term in the brackets of \eqref{eq:wderiv_fin}. 
$\frac{\partial \mathbf{F}}{\partial \mathbf{F}}$ tensor is an $M \times N$ array of $M\times N$ matrices. The $(m,n)^{th}$ matrix in the array, $\mathbf{I}_{mn}$, has all entries $\mathbf{I}_{mn}(i,j) = 0$ except for $1$ at $i=m, j=n$. 
From the definition of the tensor-vector multiplication, $\frac{\partial \mathbf{F}}{\partial \mathbf{F}}{\brho}^* = \sum_{i = 1}^M \mathbb{I}_{M \times N \times M} {\brho}^*_i$
yields an $M \times N \times M$ tensor $\tilde{\mathbb{I}} = [\tilde{\mathbf{I}}_1, \tilde{\mathbf{I}}_2, \cdots \tilde{\mathbf{I}}_M]$ where the $m$th row of $\tilde{\mathbf{I}}_m$ equals ${{\brho}^*}^T$, $0$'s otherwise. 
After another tensor-vector multiplication with conjugate error term, the second expression yields
\begin{equation}
\left(\frac{\partial \mathbf{F}}{\partial \mathbf{F}} {\brho}^*\right) (\bar{\d}^* - \bar{\d}) = (\bar{\d}^* - \bar{\d}) {\brho^*}^T.
\end{equation}
%Consider the terms that are summed over layers of the network in equation \eqref{eq:wderiv_fin}. 
Taking the first component inside the brackets of equation \eqref{eq:wderiv_fin}, we denote
\begin{equation}
\partial^k_{\mathbf{F}} \ell = \sum_{i = 1}^{L} \frac{\partial \brho^k}{\partial \mathbf{F}} \frac{\partial \brho^L}{\partial \brho^k}  \frac{\partial {\brho}^*}{\partial \brho^L}\frac{\partial \ell(\d^*, \d) }{\partial {\brho}^*}.
\end{equation}
For each partial derivative of $\brho^L$ with respect to other representations in the network, we can write the chain rule as ${\partial \brho^L}/{\partial \brho^k} = ({\partial \brho^{k+1}}/{\partial \brho^k}) ({\partial \brho^L}/{\partial \brho^{k+1}})$. 
Moving down the network beginning from layer $L$, this derivative can be evaluated by multiplying the ${\partial \brho^{k+1}}/{\partial \brho^k}$ term repeatedly for $k = L-1, L-2, \cdots 1$. 
Denote $\mathbf{y}^{k} =  | \mathbf{z}^k |$, where $\mathbf{z}^k = \mathbf{Q} \brho^{k-1} + \alpha \mathbf{F}^H \d$. %such that $\mathbf{y}_i^{k} = \sqrt{\mathbf{z}_i^k \bar{\mathbf{z}_i}^k}$. 
The partial of $\brho^k$ with respect to $\brho^{k-1}$ can be evaluated as:
\begin{equation}
\frac{\partial \brho^{k}}{\partial \brho^{k-1}} = (\frac{\partial \mathbf{z}^k}{\partial \brho^{k-1}} \frac{\partial \mathbf{y}^k}{\partial \mathbf{z}^k} + \frac{\partial \bar{\mathbf{z}}^k}{\partial \brho^{k-1}}  \frac{\partial \mathbf{y}^k}{\partial \bar{\mathbf{z}}^k} )\frac{\partial \brho^k}{\partial \mathbf{y}^k}.
\end{equation}
${\partial \brho^k}/{\partial \mathbf{y}^k}$ is merely the derivative of the thresholding function $\mathcal{P}_{\tau \ell_1}(\cdot)$, which is a diagonal matrix with entries $1$ at indexes where $\mathbf{y}^k_i> \tau$, and $0$ otherwise. 
Similarly, ${\partial \mathbf{y}^k}/{\partial \bar{\mathbf{z}}^k}$ and ${\partial \mathbf{y}^k}/{\partial \mathbf{z}^k}$ yield diagonal matrices with entries, for the $i^{th}$ diagonal term, ${\mathbf{z}_i^k}/({2 | \mathbf{z}_i^k |})$ and ${\bar{\mathbf{z}}_i^k}/({2 | \mathbf{z}_i^k |})$, respectively. 
Finally, the partial derivatives of $\bar{\mathbf{z}}^k$ and $\mathbf{z}^k$ with respect to $\brho^{k-1}$ yield $\mathbf{Q}^H$ and $\mathbf{Q}^T$ respectively. Since $\mathbf{Q}$ is Hermitian symmetric, we have:
\begin{equation}
\frac{\partial \brho^{k}}{\partial \brho^{k-1}} = \text{Re}\left( \mathbf{Q} \ \text{diag}(\frac{\mathbf{z}_i^k}{| \mathbf{z}_i^k |}) \right) \text{diag}(\mathcal{I}_{\mathbf{y}^k > \tau} ),
\end{equation}
%using that $\mathbf{Q}$ is fixed and Hermitian symmetric, 
with $I_{\mathbf{y}^k > \tau}$ denoting the set of indexes $i$ such that $\mathbf{y}_i^k > \tau$. %is satisfied, and $\text{diag}(\cdot)$ denotes a diagonal matrix obtained from the elements of the vector argument.  
For the $\mathbf{F}$ derivative of the network representations, following the same notation, we have
\begin{equation}
\frac{\partial {\brho}^k}{\partial \mathbf{F}} = (\frac{\partial \mathbf{z}^k}{\partial \mathbf{F}} \frac{\partial \mathbf{y}^k}{\partial \mathbf{z}^k} + \frac{\partial \bar{\mathbf{z}}^k}{\partial \mathbf{F}}  \frac{\partial \mathbf{y}^k}{\partial \bar{\mathbf{z}}^k} ) \frac{\partial \brho^k}{\partial \mathbf{y}^k}.
\end{equation}
%Notably, invoking the property that the Wirtinger derivative of a complex quantity with respect to its conjugate is $0$, the component from $\mathbf{z}^k$ derivative vanishes due to its dependence only on the Hermitian adjoint of $\mathbf{F}$. 
%due to Wirtinger derivative of $\mathbf{F}$ 
The first component in the parenthesis vanishes because $\mathbf{z}^k$ only depends on $\mathbf{F}^H$. {This results from the property of Wirtinger derivatives, such that $\frac{\partial \bar{c}}{\partial {c}} = 0$, for a complex variable $c \in \mathbb{C}$.}
%With $ \bar{\mathbf{z}}^k = \bar{\mathbf{Q}} \brho^k + \alpha \mathbf{F}^T \bar{\mathbf{d}}$
%we have a $M \times N \times N$ tensor, %
From $\frac{\partial \bar{\mathbf{z}}^k}{\partial \mathbf{F}} = \alpha \frac{\partial (\mathbf{F}^T \bar{\d})}{\partial \mathbf{F}}$, indexing the $3^{rd}$ dimension of the resulting $M \times N \times N$ tensor with $i$, at each $i$, with subscript $:,j$ denoting the $j^{th}$ column, we have
$$
\left(\frac{\partial (\mathbf{F}^T \bar{\d})_i}{\partial \mathbf{F}}\right)_{:, j} = \begin{cases}
\bar{\d} \ \ \text{if} \ j = i \\
\mathbf{0} \ \ \text{else}
\end{cases}. 
$$
%Following the tensor algebra and 
Denoting previous terms computed as $\partial \ell_{\brho^k}$, the $\mathbf{F}$-derivative at the layer $k$ becomes
\begin{equation}
(\partial^k_{\mathbf{F}} \ell)_{:,i} = \frac{\alpha(\partial \ell_{\brho^k})_i}{2} \frac{(\mathbf{Q} \brho^k + \alpha \mathbf{F}^H \mathbf{d})_i}{|(\mathbf{Q} \brho^k + \alpha \mathbf{F}^H \mathbf{d})_i|} \bar{\d},
\end{equation}
if $|(\mathbf{Q} \brho^k + \alpha \mathbf{F}^H \d)_i| = \mathbf{y}_i^k > \tau$ and $0$ everywhere else.

Finally, multiplying with the partial derivative of $\mathbf{F} = \text{diag} (\w) \tilde{\mathbf{F}}$, we obtain the derivative with respect to $\w$ by, for index $i = 1, 2, \cdots M$, and subscript $i,:$ denoting the $i^{th}$ $1 \times N$ row of the corresponding matrix:
\begin{equation}
(\frac{\partial \ell}{\partial \w} )_i =  \tilde{\mathbf{F}}_{i,:} \left(\frac{\partial \ell}{\partial \mathbf{F}} \right)^T_{i,:}.
\end{equation}

\subsection{Threshold Derivative}
%Since all inputs to the activation function are positive as $\brho^k_i = \max (0, \mathbf{y}^k_i - \tau)$, the derivative $({\partial \brho^k}/{\partial \tau})_{1 \times N}$ will equal $-1$ at indexes $\mathbf{y}^k_i > \tau$ and $0$ otherwise. Then, the $k$th layer derivative becomes
Since $\brho^k_i = \max (0, \mathbf{y}^k_i - \tau)$, the derivative $({\partial \brho^k}/{\partial \tau})_{1 \times N}$ will equal $-1$ at indexes $\mathbf{y}^k_i > \tau$ and $0$ otherwise. Then, the $k$th layer derivative becomes
\begin{equation}\label{eq:Tau_deriv}
\frac{\partial \ell}{\partial \tau} = \sum_{k = 1}^L \sum_{i \in \mathcal{I}_{\mathbf{y}^k > \tau}} - \left( \frac{\partial \brho^L}{\partial \brho^k} \frac{\partial {\brho}^*}{\partial \brho^L}\frac{\partial \ell(\d^*, \d) }{\partial {\brho}^*}\right)_i,
\end{equation}
where $\mathcal{I}_{\mathbf{y}^k > \tau}$ is again the set of indexes where $\mathbf{y}^k = |\mathbf{Q}_i \brho^k + \alpha \mathbf{F}^H_i \mathbf{d}|  > \tau$.

\end{document}